\documentclass{aa}
\usepackage{graphicx}
\usepackage{txfonts}
\usepackage{hyperref}
\usepackage{physics}
\usepackage{xcolor}
\usepackage{orcidlink}

\begin{document}

   \title{Large-scale radio bubbles around the black hole transient V4641~Sgr}

   \subtitle{}

   \author{N. Grollimund\inst{1}\fnmsep\thanks{E-mail: noa.grollimund@cea.fr}\orcidlink{0009-0005-4358-5146}
          \and
          S. Corbel\inst{1}\orcidlink{0000-0001-5538-5831}
          \and
          R. Fender\inst{2, 3}\orcidlink{0000-0002-5654-2744}
          \and
          J. H. Matthews\inst{2}\orcidlink{0000-0002-3493-7737}
          \and
          I. Heywood\inst{2, 4, 5}\orcidlink{0000-0001-6864-5057}
          \and
          F. J. Cowie\inst{2}\orcidlink{0009-0009-0079-2419}
          \and
          A. K. Hughes\inst{2}\orcidlink{0000-0003-0764-0687}
          \and
          F. Carotenuto\inst{6}\orcidlink{0000-0002-0426-3276}
          \and
          S. E. Motta\inst{7}\orcidlink{0000-0002-6154-5843}
          \and
          P. Woudt\inst{3}\orcidlink{0000-0002-6896-1655}
          }
    \authorrunning{N. Grollimund et al.} 

   \institute{
         Université Paris Cité, Université Paris-Saclay, CEA, CNRS, AIM, F-91191 Gif-sur-Yvette, France
         \and
         Astrophysics, Department of Physics, University of Oxford, Keble Road, Oxford OX1 3RH, UK
         \and
         Department of Astronomy, University of Cape Town, Private Bag X3, Rondebosch 7701, South Africa
         \and
         Department of Physics and Electronics, Rhodes University, PO Box 94, Makhanda 6140, South Africa
         \and
         South African Radio Astronomy Observatory, 2 Fir Street, Black River Park, Observatory 7925, South Africa
         \and
         INAF-Osservatorio Astronomico di Roma, Via Frascati 33, I-00078 Monte Porzio Catone (RM), Italy
         \and
         Istituto Nazionale di Astrofisica, Osservatorio Astronomico di Brera, via E. Bianchi 46, 23807 Merate (LC), Italy
             \\
             }

   \date{Received XX; accepted XX}
 
  \abstract 
   {Black holes (BHs) in microquasars can launch powerful relativistic jets that have the capacity to travel up to several parsecs from the compact object and interact with the interstellar medium. Recently, the detection of large-scale very-high-energy (VHE) gamma-ray emission around the black hole transient V4641~Sgr and other BH-jet systems suggested that jets from microquasars may play an important role in the production of galactic cosmic rays.}
   {V4641~Sgr is known for its superluminal radio jet discovered in 1999, but no radio counterpart of a large-scale jet has been observed. The goal of this work is to search for a radio counterpart of the extended VHE source.}
   {We observed V4641~Sgr with the MeerKAT radio telescope at the \textit{L} and \textit{UHF} bands and produced deep maps of the field using high dynamic range techniques.}
   {We report the discovery of a large-scale ($\sim 35~\mathrm{pc}$), bow-tie-shaped, diffuse, radio structure around V4641~Sgr, with similar angular size to the extended X-ray emission discovered by XRISM. However, it is not spatially coincident with the extended VHE emission. After discussing the association of the structure with V4641~Sgr, we investigate the nature of the emission mechanism. We suggest that the bow-tie structure arose from the long-term action of large-scale jets or disk winds from V4641~Sgr. If the emission mechanism is of synchrotron origin, the radio/X-ray extended structure implies acceleration of electrons up to more than $100~\mathrm{TeV}$ as far as tens of parsecs from the black hole.}
   {}


   \maketitle

\section{Introduction}

Relativistic jets are an ubiquitous facet of accretion onto compact objects, from stellar-mass black holes (BHs) to supermassive black holes (SMBHs). They release a large amount of the accreted power back to their environment, energizing the interstellar medium and influencing galactic evolution. In BH low-mass X-ray binaries (BH LMXBs), a companion star feeds a hot accretion disk surrounding the BH via Roche lobe overflow. In this configuration, the compact object is able to convert a significant fraction of the infalling material into powerful relativistic outflows on a wide range of scales. Such systems, nicknamed microquasars, are considered as smaller analogs of active galactic nuclei (AGNs).

While LMXBs spend most of their time in quiescence, they occasionally enter outburst phases during which their accretion rate and X-ray luminosity increase by several orders of magnitude. During an outburst, the system usually cycles through different accretion states defined by specific spectral and timing properties \citep{HomanBelloni2005, Remillard_McClintock2006, Belloni_Motta2016}. During transitions between accretion states, discrete ejecta, in the form of bipolar, relativistic plasma bubbles, can be launched \citep[e.g.,][]{Corbel2004, Fender2004, Miller-Jones2012}. These transient jets often display apparent superluminal motion and they have been historically observed at sub-parsec scales \citep[e.g., GRS 1915+105, GRO J1655--40;][]{Mirabel1994, Hjellming1995}. The associated radio emission is characteristic of radiation from particles accelerated at long-lived shocks \citep{Fender2023, Cooper2025, Matthews2025, Savard2025}. For some sources, these transient jets can persist out to parsec-scales when they interact with the interstellar medium (ISM). Furthermore, recurrent outbursts (and thus jet production) have been shown to produce parsec-scale lobes, as seen in the case of 1E 1740.7--2942 \citep{Mirabel1992} and GRS 1758-258 \citep{MirabelRodriguez1999, Marti2017}. Emission from large-scale discrete ejecta has also been detected up to X-rays in several systems, such as XTE J1550--564 \citep{Corbel2002, Tomsick2003, Kaaret2003, Migliori2017}, H 1743--322 \citep{Corbel2005} and MAXI J1820+070 \citep{Espinasse2020}, implying particle acceleration up to $\sim10~\mathrm{TeV}$. Over the last few years, observations carried out with the MeerKAT and ATCA arrays have revealed the omnipresence of large-scale jets from BH LMXB, including MAXI J1535--571 \citep{Russell2019}, MAXI J1820+070 \citep{Bright2020, Espinasse2020}, MAXI J1348--630 \citep{Carotenuto2021a}, MAXI J1848--015 \citep{Bahramian2023}, and 4U 1543--47 \citep{Zhang2025}, hinting that all BH LMXBs are probably microquasars.

Jets appear to be essential in the dynamics of the overall accretion flow in BH systems and are thought to be the main channel for the transfer of accretion energy to the surrounding environment \citep{FenderMunoz-Darias2016}. Recently, the detection of very-high-energy (VHE) gamma-ray emission associated with BH-jet systems demonstrated that stellar-mass BHs and their environments can operate as extremely efficient accelerators of particles, suggesting that large-scale jets from BHs could be a notable source of galactic cosmic rays \citep{Alfaro2024, LHAASO2025}, as previously proposed by \citet{FenderMaccaronevanKesteren2005} and \citet{Cooper2020}. Among these microquasars, the black hole X-ray binary V4641~Sgr was the brightest (by an order of magnitude) and displayed extended gamma-ray emission with one of the hardest spectra, continuing up to 0.8 PeV.

V4641 Sagitarii (V4641~Sgr) is a low-mass\footnote{Even though the companion is a B-type star, \citet{MacDonald2011} argue the maintenance of the LMXB classification because mass transfer occurs via Roche lobe overflow.} X-ray binary hosting a black hole with a mass of $6.4 \pm 0.6~M_\odot$ and a B9III stellar companion with a mass of $2.9 \pm 0.4~M_\odot$ \citep{MacDonald2014}. This transient, located at a parallax distance of $6.2 \pm 0.7~\mathrm{kpc}$ \citep{MacDonald2014, Gandhi2019} was discovered in 1999 by \textit{BeppoSAX} \citep{intZand1999} and \textit{RXTE} \citep{Markwardt1999} when it exhibited a bright X-ray flare, although an outburst in 1978 was found on photographic plates decades later \citep{Barsukova2014}. A radio counterpart was identified by \citet{Hjellming2000} and revealed short-lived relativistic jets. The highly superluminal motion ($\sim 10c$) of these discrete ejecta \citep[which is notably a matter of debate; e.g., ][]{Marti2026} would imply a low jet inclination angle ($\la10\degr$), while strong ellipsoidal variations were found to indicate a high orbital axis inclination angle \citep[$\ga~60\degr$;][]{Orosz2001}, suggesting a strong spin-orbit misalignment \citep{Maccarone2002, Salvesen2020}. Since 1999, V4641~Sgr has been observed in outburst every one to three years, always exhibiting short bursting activity. In September 2024, a new outburst was detected by MAXI \citep{Negoro2024ATel}, while a small radio flare was observed by MeerKAT a month later \citep{Grollimund2024ATel}.

The detection of a 1--2$\degr$ jet-like gamma-ray structure around V4641~Sgr by HAWC \citep{Alfaro2024}, LHAASO \citep{LHAASO2025}, and H.E.S.S. \citep{HESS2025} suggests that particle acceleration beyond 1 PeV (if hadronic) takes place at similar distances from the black hole as the well-known microquasar SS 433 \citep{Abeysekara2018, HESS2024}. Furthermore, recent X-ray observations with XRISM indicate the presence of another extended component located closer to V4641~Sgr \citep{Suzuki2025}. The spatial extent of this structure radius of $7' \pm 3'$) is much smaller than the size of the gamma-ray bubble, hinting that the X-rays are produced by a population of electrons closer to the black hole than the particle acceleration sites responsible for the gamma-ray emission.

In this work, we report on the discovery of a large-scale ($20'$ in diameter, i.e., $35~\mathrm{pc}$) radio structure around V4641~Sgr in deep MeerKAT observations. This paper is structured as follows. Section~\ref{sec:obs} details the observations and data reduction we carried out, Sect.~\ref{sec:results} presents the structure we found in the final radio map, Sect.~\ref{sec:discussion} offers a discussion of the emission mechanism, energetics, and the possible association of this structure with large-scale jets or disk winds. Section~\ref{sec:conclusions}  presents out conclusions.

\section{Observations and data reduction}
\label{sec:obs}

\begin{figure*}
    \centering
    \includegraphics[width=\linewidth]{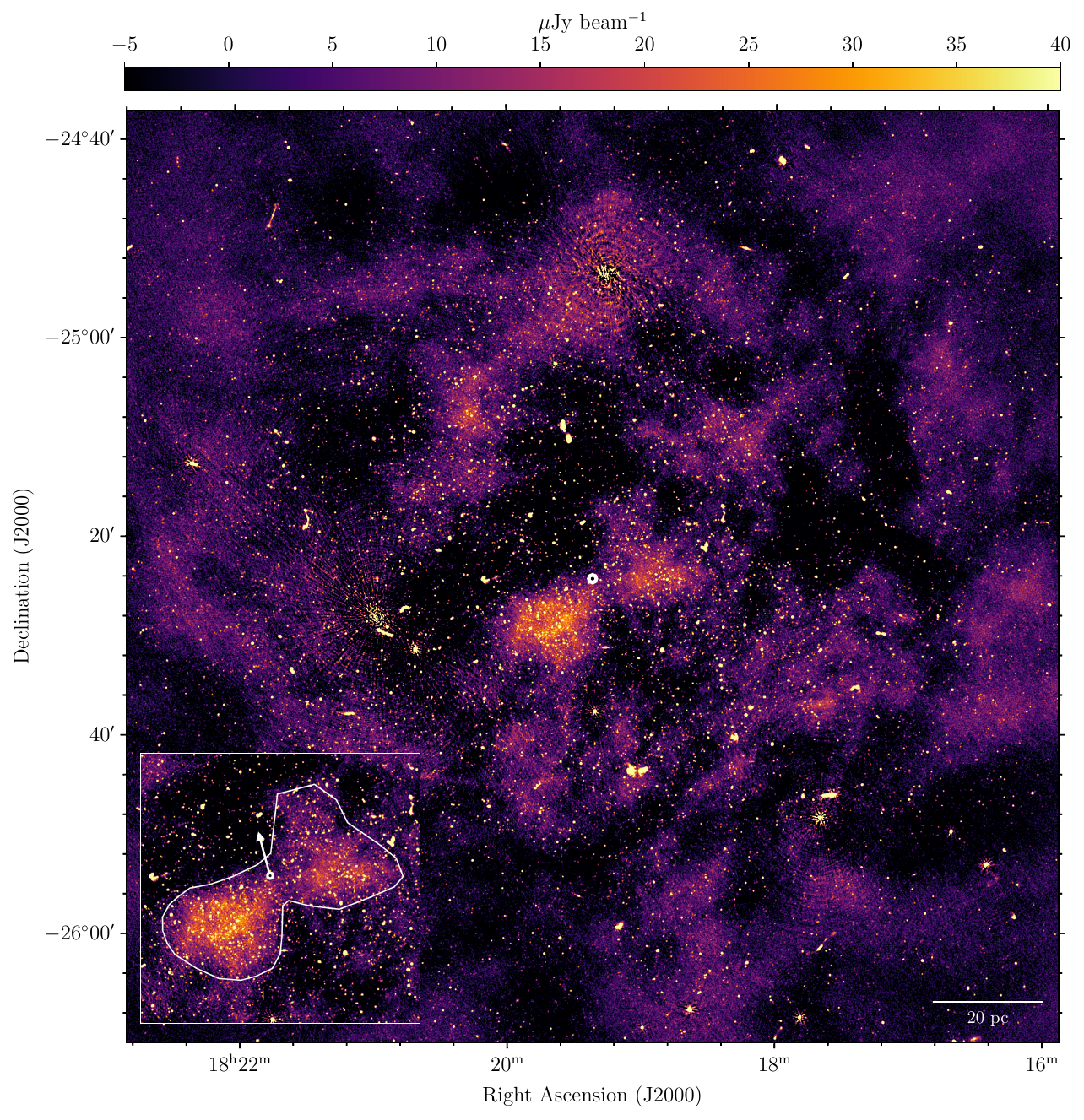}
    \caption{V4641~Sgr field as observed by the MeerKAT interferometer in the \textit{L} band (central frequency of 1.217 GHz), with an angular resolution of $8''$. The background noise level is $4~\mu\mathrm{Jy~beam^{-1}}$, while the noise level in the region surrounding the bow-tie is on the order of $10~\mu\mathrm{Jy~beam^{-1}}$. The position of V4641~Sgr is marked by a black dot inside a white circle. The inset shows the central region of the field, with the bow-tie outlined in white. The white arrow indicates the proper motion of V4641~Sgr with respect to nearby stars, in the north-northeastern (NNE) direction (position angle of $15\degr$ E of N; see Sect. \ref{sec:proper_motion}).}
    \label{fig:V4641_map}
\end{figure*}

We observed V4641~Sgr in radio with the MeerKAT telescope from 2020 to 2024 as part of the ThunderKAT Large Survey Project \citep{Fender2018} and X-KAT large MeerKAT open-time program. Most of the observing runs took place on a weekly cadence, during the 2021-2022 and 2024 outbursts (see Table \ref{tab:obs_meerkat}), using \textit{L}-band receivers (856–1712 MHz). We also performed a 1-hr \textit{UHF}-band (544-1088 MHz) observation of the source. The correlator was configured to deliver either 4096 or 32768 channels across the total bandwidth, which were binned down to 1024 channels, with an 8-second integration time per visibility point. We used PKS J1939--6342 as a primary calibrator to set the absolute flux and bandpass, and J1833--2103 as a phase calibrator to perform a complex gain calibration. Each run consisted of a single 15-min scan or several 30-min scans of V4641~Sgr, interleaved by 2-min scans of the phase calibrator, as well as a single 10-min scan of the primary calibrator.

We reduced and imaged each of the 32 observations using the \textsc{oxkat\footnote{\url{https://github.com/IanHeywood/oxkat}}} data reduction scripts \citep{oxkat}. The visibilities were initially flagged and calibrated using the Common Astronomy Software Application package \citep[\textsc{casa};][]{casa}. Additional flagging on the target data was conducted using the \textsc{tricolour}\footnote{\url{https://github.com/ratt-ru/tricolour}} package \citep{tricolour}. All imaging was done using \textsc{wsclean} \citep{wsclean} with a Briggs weighting scheme (robust parameter of 0.2 to maximize sensitivity). We generated full-band, multifrequency synthesis (MFS) images by deconvolving in eight sub-bands and convolving the resulting model image with a $8''$ circular gaussian. After a step of direction-independent self-calibration with the \textsc{cubical} package \citep{cubical}, the target was imaged a second time using masked deconvolution.

When imaging the square-degree MeerKAT field, the presence of two bright ($\sim$ 0.3--1 Jy) sources\footnote{QSO J1820--2528 ($\text{RA} = 18^{\text{h}}20^{\text{m}}57.8^{\text{s}}$, $\text{Dec} = -25\degr 28' 12.6''$) and PMN J1819--2453 ($\text{RA} = 18^{\text{h}} 19^{\text{m}} 15.1^{\text{s}}$, $\text{Dec} = -24\degr 53' 45.9''$)} resulted in severe artifacts. Thus, we  peeled these sources from each observation to improve the dynamic range of the final map. To this end, we generated a model of the field with higher angular and frequency resolution using \textsc{wsclean} and partitioned it into another FITS cube containing only the bright sources. We then predicted model visibilities from this cube, which we \textit{uv}-subtracted from the model visibilities of the full sky using \textsc{cubical}. We imaged the longest observation (4h on-source) and used the resulting map to generate a refined cleaning mask. After the peeling stage, the \textit{L}-band data were combined and imaged using \textsc{wsclean}, reaching a total exposure of $\sim 11\mathrm{h}$. The residual artifacts were further mitigated by removing 160 of the 1024 spectral channels, as we noticed that these mainly originated from the upper edge of the band. The final \textit{L}-band (central frequency of 1.217 GHz) and \textit{UHF}-band (816 MHz) continuum images were then analyzed using the Cube Analysis and Rendering Tool for Astronomy (\textsc{carta}).

\section{Results}
\label{sec:results}

\subsection{ Large-scale bow-tie structure}

The \textit{L}-band MeerKAT wide-field image of V4641~Sgr is presented in Fig. \ref{fig:V4641_map}. In this map, a large-scale, faint, bow-tie-shaped structure resides close to the pointing center. This extended emission, which is roughly symmetric and spreads across $\sim 10'$ NW and SE of V4641~Sgr, was not previously known. The low surface brightness ($\la 50~\mathrm{\mu Jy~beam^{-1}}$) and angular scale of the diffuse emission make a detection challenging. MeerKAT was able to detect the structure thanks to its excellent \textit{uv}-coverage\footnote{MeerKAT is sensitive to structures up to a maximum angular scale of $\sim 15'$ at the \textit{L} band and $\sim 25'$ at the \textit{UHF} band.} and sensitivity, while most radio interferometers would likely resolve it out. In the \textit{UHF} band, the bow-tie can also be detected and it has a very similar shape, providing evidence that the structure is real and does not arise from imaging artifacts.

To estimate the integrated flux density of the bow-tie structure, we first subtracted the point-like sources in the image plane. Indeed, the V4641~Sgr field is dominated by compact sources, including background and foreground sources which would erroneously increase the flux measurement if not subtracted. We then computed the flux density of the extended feature by integrating the brightness over two ellipsoidal regions (one for each side of the bow-tie) and obtained $S_\nu = 87 \pm 13~\mathrm{mJy}$ at $1.217~\mathrm{GHz}$ and $S_\nu = 75 \pm 26~\mathrm{mJy}$ at $816~\mathrm{MHz}$, where the uncertainties are mostly due to the choice of extraction regions. This yields a \textit{UHF/L}-band spectral index of $\alpha_r = 0.37\pm 0.95$. We note that the systematic subtraction of the point sources might be too drastic; for instance,  if some of the compact sources were actually related to the extended emission. In such cases, the flux values we obtained would be underestimated.

Interestingly, the structure we discovered is spatially coincident (same position and angular size) with the extended X-ray emission around V4641~Sgr reported by \citet{Suzuki2025}, who found a radius of $7' \pm 3'$. This is especially relevant given that this value was extracted from a one-dimensional (1D) radial profile, while an angular distance of $10'$ from the black hole would align with the edges of the bow-tie.

\begin{figure}
    \centering
    \includegraphics[width=\linewidth]{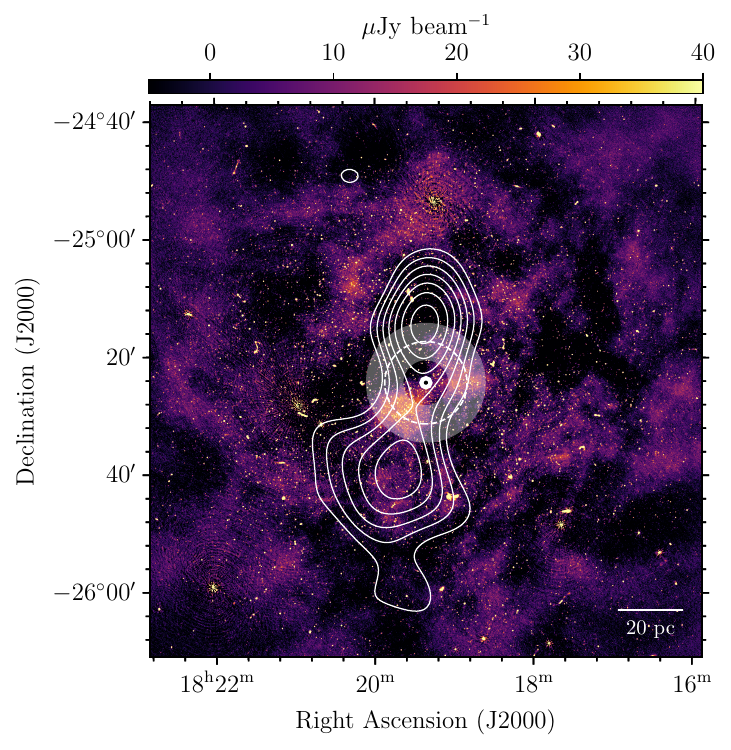}
    \caption{ V4641~Sgr field as observed by the MeerKAT interferometer in the \textit{L} band, along with H.E.S.S. gamma-ray contours (energy range of 0.8 to 22 TeV) at the levels of 3, 4, ..., 9 $\sigma$ \citep{HESS2025}. The dashed circle and the shaded area represent the radius of the XRISM X-ray source and its uncertainty \citep{Suzuki2025}.}
    \label{fig:map_hess_xrism}
\end{figure}

In Fig. \ref{fig:map_hess_xrism}, we show the MeerKAT map with the H.E.S.S. (0.8--22 TeV) contours overlaid. A similar figure with HAWC contours can be found in Appendix \ref{appendix:hess_hawc}. We did not find any clear radio counterpart to the extended gamma-ray sources reported by \citet{Alfaro2024}, \citet{LHAASO2025} and \citet{HESS2025}. The angular size of the radio structure ($\sim 20'$) is smaller than the gamma-ray jet-like features ($\sim 1\degr$) and its orientation in the plane of the sky is different (position angle of $\sim -50\degr$, while the gamma-ray bubble is roughly aligned with the N-S direction). When considering a model with two point sources, \citet{Alfaro2024} found a northern and a southern component, with angular separations to the black hole of $0.23\degr$ and $0.46\degr$, respectively. Interestingly, the positions of these components correspond to "radio cavities," namely, regions where the radio map is particularly free of diffuse emission (see Fig. \ref{fig:hess_hawc}). It is also worth noting that a faint, arc-shaped region lies directly NE of the northern gamma-ray bubble, as seen in Fig. \ref{fig:map_hess_xrism} and \ref{fig:hess_hawc}. Although we cannot confirm a direct association with the structures detected by H.E.S.S. and HAWC, it remains plausible that this region traces an edge-brightened bow shock around the gamma-ray bubble.

\subsection{Proper motions of stars in the vicinity of V4641~Sgr}
\label{sec:proper_motion}

The position, parallax, and proper motion of $\sim 1.46$ billion optical sources have been measured by the \textit{Gaia} space observatory. In this work, we  rely on \textit{Gaia} astrometric data to study the motion of V4641~Sgr with respect to stars in its vicinity. The underlying purpose of this is to discuss the association of the bow-tie with V4641~Sgr (see Sect. \ref{sec:association}).

\begin{figure}
    \centering
    \includegraphics[width=\linewidth]{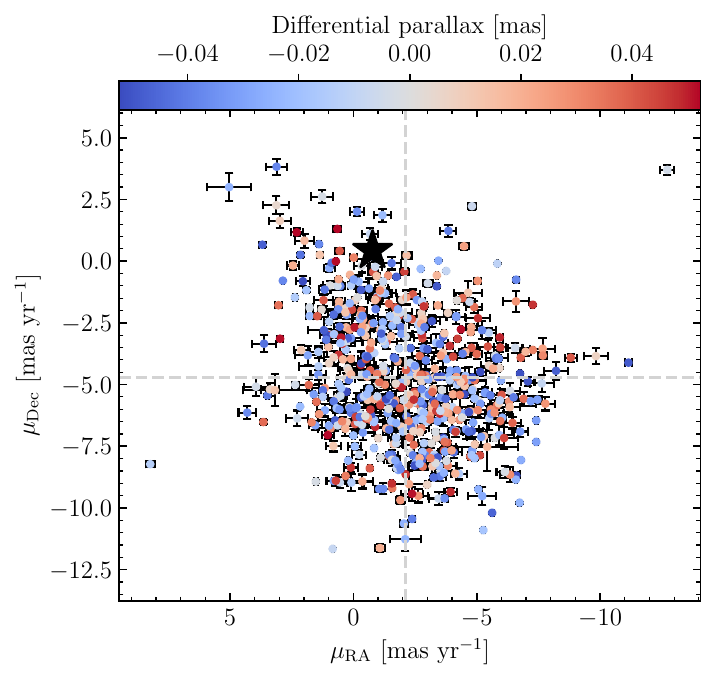}
    \caption{Proper motion of stars in the vicinity of V4641~Sgr, in the direction of increasing right ascension (\textit{x}-axis) and declination (\textit{y}-axis), as measured by \textit{Gaia}. Only stars within a $3'$ radius and a $p_\star \pm 2\sigma_{p_\star}$ parallax range are shown. The differential parallax between V4641~Sgr and individual stars is color-coded. The black star $\star$ corresponds to the optical counterpart of V4641~Sgr, while the dotted gray lines indicate the average proper motion in the sample.}
    \label{fig:propermotion}
\end{figure}

We queried the \textit{Gaia} DR3 catalog using the \textsc{astroquery\footnote{\url{https://astroquery.readthedocs.io/en/latest/}}} Python package. Nearby stars were selected through a cone search centered on V4641~Sgr, followed by the exclusion of sources with parallaxes falling outside a defined narrow interval. In Fig. \ref{fig:propermotion}, we show the proper motion of stars within $3'$ around V4641~Sgr, with parallaxes in the range $p_\star \pm 2\sigma_{p_\star}$, where $p_\star = 0.169~\mathrm{mas}$ is the V4641~Sgr parallax measurement and $\sigma_{p_\star} = 0.026~\mathrm{mas}$ is its uncertainty. The population is represented in the $(\mu_\mathrm{RA},~\mu_\mathrm{Dec})$ plane, where $\mu_\mathrm{RA}$ and $\mu_\mathrm{Dec}$ are the proper motions in the direction of increasing right ascension and declination, respectively. Another representation of the proper motion vectors is provided in Appendix \ref{appendix:propermotion}. The average proper motion within the sample is $\overline{\mu_\mathrm{RA}} = -2.11~\mathrm{mas~yr^{-1}}$, $\overline{\mu_\mathrm{Dec}} = -4.70~\mathrm{mas~yr^{-1}}$ (dotted gray lines in Fig. \ref{fig:propermotion}), implying that this population of stars has a significant global motion toward the south-west. On the other hand, V4641~Sgr displays a rather slow motion in the plane of the sky ($\mu_\mathrm{RA,~\star} = -0.78 \pm 0.03~\mathrm{mas~yr^{-1}}$, $\mu_\mathrm{Dec,~\star} = 0.43 \pm 0.02~\mathrm{mas~yr^{-1}}$, position angle of $-61\degr$ E of N). As a result, its motion relative to nearby stars is $\Delta\mu_\star = 5.31 \pm 0.04~\mathrm{mas~yr^{-1}}$ toward the north-northeast (position angle of $15\degr$ E of N). We performed the same analysis with other selection radii ($1'$, $5'$, $10'$) and parallax intervals ($p_\star \pm \sigma_{p_\star},~\pm 3\sigma_{p_\star}$) and found very similar results within uncertainties.

\section{Discussion}
\label{sec:discussion}

\subsection{Association with V4641~Sgr}
\label{sec:association}

In our MeerKAT image, V4641~Sgr appears to be located close to the center of symmetry of the bow-tie structure in the plane of the sky. In addition, the angular size of each lobe ($\sim10'$) of this large-scale feature is similar to the radius of the X-ray extended emission ($7 \pm 3'$) inferred by \citet{Suzuki2025}, hinting that the radio and X-ray structures may be related. We note that V4641~Sgr is slightly offset ($\la 1'$ towards NNE) from the apparent center of symmetry of the bow-tie. However, this could be explained by a proper motion effect: if the source has moved since it began producing the extended structure, it might no longer be positioned at the center of it. As discussed in Sect. \ref{sec:proper_motion}, V4641~Sgr is moving in the NNE direction relative to the nearby stars, and hence, to a first approximation, similarly with respect to the surrounding ISM. This is precisely the direction of the offset between the XRB and the apparent center of the large-scale structure. As the bow-tie should move with the ISM, we conclude that a proper motion effect is a viable explanation for the offset. Then, if we trace the trajectory backwards in time, we find that V4641~Sgr was at the center of the bow-tie (i.e., $\sim 1'$ away from its current position) about $10~\mathrm{kyrs}$ ago. Alternatively, we could raise the possibility of inhomogeneities of the ISM density around the source, which would lead to variations in the radiative efficiency, which would, in turn, cause the apparent center of symmetry to be shifted towards the direction where the medium density is higher. Finally, the position angle of the 1999 radio jets \citep[$\sim -18\degr$;][]{Hjellming2000} falls within the angular range of the bow-tie, since the latter has a mean axis tilted by $\sim-50\degr$ E of N, along with an apparent opening angle of $\varphi \simeq 80\degr$. We note that the mean direction from V4641~Sgr to the brightest parts of the radio lobes ($\sim -70\degr$) is significantly offset from the direction of the 1999 jets and this is a caveat.
Thus, we assume hereafter that the radio and X-ray extended emission originate from the same large-scale structure, which we associate with past activity from V4641~Sgr.

\subsection{ Jet and wind scenarios}
\label{sec:jet_scenario}

Large-scale collimated jets can generate persistent jet-like features or at least occasionally when interacting with the ISM, hotspots, and bow-shock regions. The bow-tie feature around V4641~Sgr is a large-scale, symmetric, diffuse structure, which is reminiscent of the radio lobes observed in other sources. If we assume that the radio and X-ray emission (which have similar spatial extent) are related, the global radio-to-X-rays spectral index $\alpha = -0.58\pm 0.02$ is consistent with optically thin synchrotron emission from a plasma of relativistic electrons (see Sect. \ref{sec:emission_mechanism} for a discussion on the emission mechanism). Therefore, we suggest that the bow-tie structure is the result of the long-term action of large-scale jets from V4641~Sgr.

We did not identify any clear signature of an edge-brightened bow-shock that would correspond to the compressed material at the boundaries of a jet-blown cavity, as seen in Cyg X--1, GRS 1915+105, and GRS 1758--258 \citep[][]{Gallo2005, Atri2025, Motta2025, Marti2017, Mariani2025}, or a supernova remnant around the X-ray binary, as seen in SS 433 and Cir~X--1 \citep{ClarkMurdin1978, ClarkParkinsonCaswell1975, Heinz2013}. Instead, we detected almost uniform, diffuse, extended emission around the source, which resembles the persistent jets in 4U 1755--33 \citep{Angelini2003, Kaaret2006}, although we note that the latter have a more linear shape. Such emission, if it is indeed related to a large-scale jet, would imply efficient particle acceleration all along the jet, rather than at a jet termination shock. This is reminiscent of the morphology of FR I radio galaxies, which show gradual energy dissipation and particle acceleration along the outflow axis.

The conical shape of the structure is interesting in itself. The opening angle of the bow-tie, $\psi$, is given by $\tan(\psi/2) = \tan(\varphi/2) \sin\theta$, where $\varphi \simeq 80\degr$ is the opening angle projected on the plane of the sky and $\theta$ is the angle between the symmetry axis of the bow-tie and the line of sight (i.e., the jet average inclination). Transient jets from X-ray binaries typically have measured opening angles of a few degrees \citep[see, e.g.,][]{MillerJones2006}. If $\theta$ is small, the real opening angle $\psi$ is likely much smaller than $80\degr$. If $\theta$ is larger, the observed structure might not be explained by a steady, fixed-axis outflow; instead, a precessing jet with a wide precession angle could, in principle, reproduce the bow-tie. Such a phenomenon has been observed in several microquasars, one of the best examples being the mas-scale precessing jets of SS 433 \citep[e.g.,][]{Mioduszewski2003}. Likewise, relativistic precessing jets in Cir~X--1 have been inferred to explain the powering of large diffuse structures \citep{Sell2010, Cowie2025, Cowie2025subm} with a remarkably similar morphology (including the opening angle and physical size). The possibility of a precessing accretion flow around V4641~Sgr was already suggested by \citet{Gallo2014}, who found rapid time-scale variability of a broad emission feature in the X-ray spectrum that could be explained by a variable "microblazar" behavior \citep[see also][]{Orosz2001, Chaty2003}. 

As highlighted in Sect. \ref{sec:results}, the VHE gamma-ray bubble and bow-tie structure discovered by MeerKAT have different orientations in the plane of the sky (position angles of $\sim0\degr$ and $\sim-50\degr$ E of N, respectively). Additionally, a discrepancy in terms of angle to the line of sight has been suggested by \citet{Alfaro2024}, who argued that the gamma-ray sources would be consistent with jets being perpendicular to the accretion disk \citep[with an inclination angle $i = 72.3 \pm 4.1\degr$;][]{MacDonald2014}, while the superluminal radio jet may be almost aligned with the line of sight \citep[$< 16\degr$;][]{Salvesen2020}. It has been proposed that the misalignment of the radio jet and the VHE extended structure could be explained by assuming that the latter is not directly linked to the jet. Instead, the gamma-ray feature could arise from high energy particles escaping at the jet termination (where the jet loses its stability) and streaming along the ordered magnetic field lines in the ISM \citep{Neronov2025}. In the jet scenario, the radio and X-ray extended emission is due to relativistic electrons producing synchrotron radiation. These electrons would cool quite quickly relative to protons, explaining why the low-energy structures have a lower spatial extent ($\sim 20'$) than the gamma-ray bubble ($\ga 1\degr$), which is probably of hadronic origin. Hence, the bow-tie would be caused by recent jet activity, while the VHE features would trace longer timescale energy input into the ISM.

Another possibility that could explain the formation of the bow-tie structure is related to disk winds. Optical spectroscopic studies revealed that V4641~Sgr produces strong wind outflows, detected during several outbursts \citep[see, e.g.,][]{Chaty2003, Munoz-Darias2018}. In particular, observations taken after the 1999 peak flux implied outflow velocities of up to $\sim 3000~\mathrm{km~s^{-1}}$. These disk winds, which have been observed both in soft and hard states \citep[i.e., even when the jet is active;][]{Munoz-Darias2018}, can drive blast waves propagating through the surrounding medium, accelerating particles at the shock front. In the case of radio-quiet AGNs, it has been shown that synchrotron radiation from wind-driven shocks can be a strong source of radio emission \citep{Stocke1992, Nims2015, Rankine2021, Richards2021}. In LMXBs (particularly in a strong wind source such as V4641~Sgr) a similar phenomenon could be at play. The orientation of the disk winds is likely to differ from the direction of the jet, which could explain the discrepancy between the axes of the bow-tie and the gamma-ray bubble, without the need for a precessing jet. While the inclination of the system ($72\degr$), angle to the line of sight, and position angle of the superluminal jet ($< 16\degr$ and $-18\degr$ E of N) are known, we cannot constrain the angle of the disk normal in the plane of the sky due to the strong spin-orbit misalignment \citep{Orosz2001, MacDonald2014}. For an equatorial disk wind along the bow-tie axis, this angle would be on the order of $40\degr$ E of N.

\subsection{Emission mechanism}
\label{sec:emission_mechanism}

Another key question is the nature of the emission mechanism. We plot in Fig. \ref{fig:sed} the broadband spectral energy distribution (SED) of the extended emission around V4641~Sgr. While ultra-high-energy gamma-rays have been detected by H.E.S.S., HAWC, and LHAASO up to $\sim 1\degr$ from the X-ray binary, it is known that the X-ray and radio emission regions have a spatial extent of 7--$10'$. Thus, we can make the assumption that there exist two emission regions, which are not co-spatial and which are responsible for different parts of the SED. In this scenario, the two regions are associated with two populations of particles: hadrons at large distances from V4641~Sgr and leptons, closer to the black hole. In contrast, we note that a leptonic interpretation of the gamma-ray emission was proposed by \citet{Wan2025arXiv} and \citet{HESS2025}, but we do not discuss this notion here.

Due to significant uncertainties on the flux measurements, the radio spectral index is not well constrained ($\alpha_r = 0.37\pm 0.95$): within the error bars, it is consistent with both optically thin free-free (flat spectrum) and optically thin synchrotron radiation ($\alpha~<~0$). If, instead, we include both the radio and X-rays, we obtain a spectral index of $\alpha = -0.58\pm 0.02$, which is typically expected in the case of synchrotron emission from an optically thin plasma and is only $1\sigma$ from $\alpha_r$.

\subsubsection{Synchrotron}
\label{sec:synchrotron}

To test the synchrotron hypothesis, we tentatively performed SED fitting using the \textsc{naima}\footnote{\url{http://github.com/zblz/naima.git}} package \citep{naima} with the following model:

\begin{itemize}
    \item a population of relativistic electrons radiates through synchrotron and inverse Compton (IC) scattering of soft photons;
    \item a population of relativistic protons produces gamma-rays through p-p interactions followed by pion decay;
    \item each of these populations follows a power-law distribution with an exponential cut-off: 
    \begin{equation*}
        \dv*{N}{E} = k (E/E_0)^{-p} \exp(-E/E_\mathrm{c}),
    \end{equation*}
    where $p$ is the distribution index, $E_0$ the reference point energy, $E_\mathrm{c}$ the cut-off energy, and $k$ a normalization factor.
\end{itemize}

\begin{figure}
    \centering
    \includegraphics[width=\linewidth]{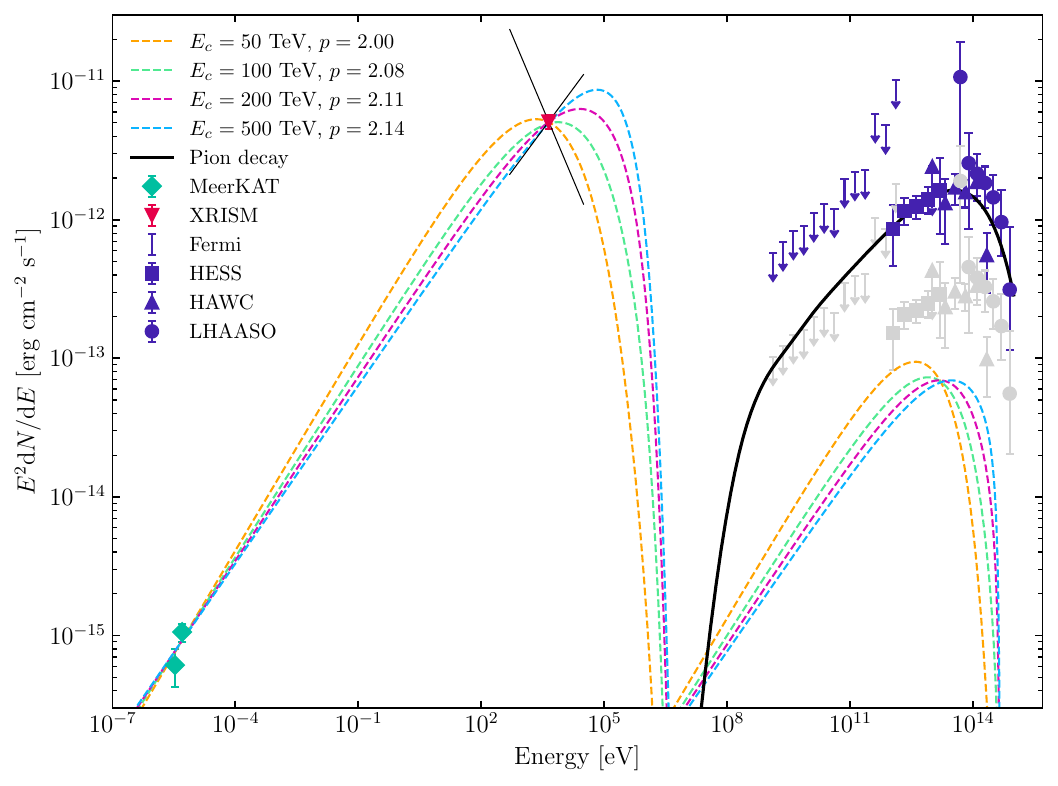}
    \caption{Broadband spectral energy distribution. Emission from the central $10'$ around V4641~Sgr is shown (coincident MeerKAT and XRISM detections), together with \textit{Fermi}-LAT, H.E.S.S., HAWC and LHAASO spectra \citep{Suzuki2025, Neronov2025, HESS2025, Alfaro2024, LHAASO2025} at larger scales ($\sim 1\degr$). The gray points correspond to gamma-ray fluxes, scaled down to the XRISM source region by relative area. Synchrotron and IC models (dotted lines), as well as the pion decay model (black solid line) are overlaid.}
    \label{fig:sed}
\end{figure}

We first fitted a synchrotron model to the radio and X-ray fluxes from the extended sources, inferring an electron distribution. Since the available data are insufficient to constrain the exponential cut-off, we fixed the value of the magnetic field to $B = 20~\mu\mathrm{G}$ \citep[an average value from ][]{Suzuki2025} and fitted the model for different cut-off energies, $E_c \in [50,\,100,\,200,\,500]~\mathrm{TeV}$. We then computed the expected IC flux from the same region assuming the inferred electron distribution. We finally fitted a pion decay model to the gamma-ray fluxes (larger region), inferring a proton distribution. The results are shown in Fig. \ref{fig:sed}. The set of parameters ($B,~E_c$) we have examined allows to reproduce the observed radio and X-ray fluxes with a synchrotron model. We find an electron distribution index of $p\sim2.1$, in agreement with the value of $\alpha$ determined above, as $\alpha = (1-p)/2 \simeq -0.55$ in the case of optically thin synchrotron radiation. Assuming this electron distribution, the computed IC flux is lower than the measured gamma-ray flux, even when the latter is scaled down to the radio/X-ray source region by relative area (gray points), as shown in Fig. \ref{fig:sed}. Under our assumptions, this is consistent with the need for a hadronic mechanism for the VHE emission in the central $10'$ (i.e., where radio, X-ray, and VHE emissions are all detected). We note that the inferred electron distribution (and, thus, the IC flux) depends on the magnetic field, which we have fixed to $B = 20~\mu\mathrm{G}$. A caveat is that a low magnetic field would imply a much higher IC flux. In that case, determining the VHE emission mechanism (which is beyond the scope of this paper) is actually not trivial. Indeed, a sharp increase in IC flux would imply a leptonic contribution to the VHE emission. For illustrative purposes, we assume a hadronic origin. Fitting the pion decay model to the gamma-ray data gives a proton distribution index of $p = 1.73_{-0.15}^{0.12}$ and a cut-off energy of $E_c = 2.5^{+4.2}_{-1.1}~\mathrm{PeV}$, consistent with the PeVatron nature of the source, as suggested by previous studies \citep{Alfaro2024, LHAASO2025, Neronov2025}.

Let us assume that the radio/X-ray spectrum arises from optically thin synchrotron emission from a relativistic plasma of electrons. We can derive a minimum energy associated with this plasma, which occurs close to equipartition between energies in electrons and magnetic field \citep[see ][]{Longair2011}. A detailed view on the computation of the minimum energy and the equipartition field is provided in Appendix \ref{appendix:min_energy}. We used a frequency range of $\nu_1 = 1.2\times 10^9~\mathrm{Hz}$ to $\nu_2 = 2.4\times 10^{18}~\mathrm{Hz}$ with a spectral index of $\alpha = (1-p)/2 = -0.6$ and a flux density of $S_{\nu_0} = 45.4~\mathrm{nJy}$ at the pivot frequency of $\nu_0 = 4.2\times 10^{17}~\mathrm{Hz}$. This gives an integrated luminosity of $8.1\times 10^{34}~\mathrm{erg~s^{-1}}$.

\begin{figure}
    \centering
    \includegraphics[width=\linewidth]{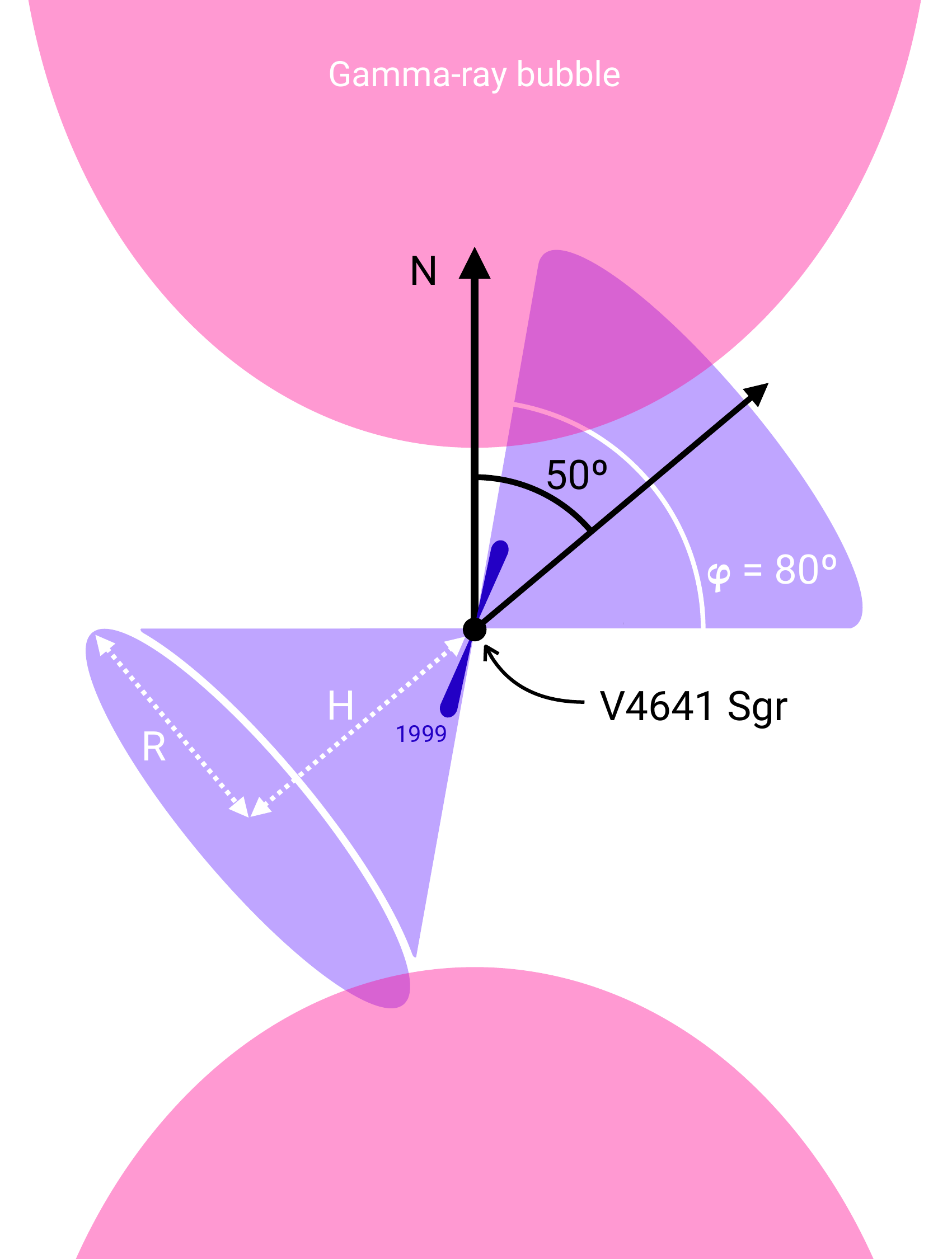}
    \caption{Schematic illustration of the bow-tie around V4641~Sgr in the plane of the sky. In our model, the geometry of the bow-tie is a cone of radius, $R$, and height, $H$, with V4641~Sgr at its vertex. The symmetry axis of the cone has a position angle of $-50\degr$ E of N. $\varphi$ is the projected opening angle of the structure. The black dot corresponds to the position of V4641~Sgr, while the direction of the 1999 radio jets \citep[$-18\degr$ E of N;][]{Hjellming2000} is represented in dark blue. The pink shaded areas symbolize the gamma-ray bubbles \citep{Alfaro2024, LHAASO2025, HESS2025}.}
    \label{fig:bowtie_schematic}
\end{figure}

To estimate the volume of the bow-tie structure, we assumed a conical geometry, as shown in Fig. \ref{fig:bowtie_schematic}. This is motivated by the shape of the radio structure and the possible association of the latter with a large-scale jet (see Sect. \ref{sec:jet_scenario} for a discussion). When placing V4641~Sgr at the vertex of a cone, we measure a height of $10'$ and a radius of $6'$, which corresponds (at 6.2 kpc) to physical scales of $H \simeq 18(\sin\theta)^{-1}~\mathrm{pc}$ and $R \simeq 11~\mathrm{pc}$, respectively, and a total volume of $V = 1.3 \times 10^{59}(\sin\theta)^{-1}~\mathrm{cm}^3$. Finally, we assume $\eta = 1$ (see Appendix \ref{appendix:min_energy}) for consistency with similar studies, and a filling factor $f = 0.5$. We find an equipartition field of $B_\mathrm{eq} = 5.8(\sin\theta)^{2/7}~\mu\mathrm{G}$ and a corresponding minimum total energy of $E_\mathrm{min} = 2.1\times 10^{47}(\sin\theta)^{-3/7}~\mathrm{erg}$. To produce the observed X-ray emission, electrons in a magnetic field $B_\mathrm{eq}$ need a Lorentz factor of $\gamma = 2.6 \times 10^8$. Thus, electrons within the large-scale bow-tie structure are accelerated up to $130~\mathrm{TeV}$ and have a cooling timescale of $2.9~\mathrm{kyrs}$ in the absence of adiabatic cooling. This also leads to an estimate for the number of electrons of $7.5 \times 10^{48}$ and, assuming a proton-electron plasma, a total mass of $1.3 \times 10^{25}~\mathrm{g}$. The above physical quantities also depend on $\sin\theta$ to a negative power and are thus lower limits.

To check whether a jet scenario is realistic, we compare tentative estimations of the jet age and accretion timescales. A lower limit on the age of the structure can be determined by assuming that the bow-tie expands at $10\%$ of the speed of light as a typical terminal speed measured in other systems \citep[e.g., Sco X--1, XTE J1550–564;][]{Fomalont2001a, Fomalont2001b, Migliori2017}, keeping in mind the true expansion velocity is likely smaller. For an angular size of $10'$ (i.e., $\sim 18(\sin\theta)^{-1}~\mathrm{pc}$ at $D = 6.2~\mathrm{kpc}$) on either side of V4641~Sgr, this gives a minimum jet age of $\sim 590(\sin\theta)^{-1}~\mathrm{yrs}$. A more reasonable value would be the synchrotron cooling time of the high energy electrons (i.e., 2.9 kyrs) derived above using synchrotron minimum energy arguments. Lastly, the travel time of V4641~Sgr, from the center of the bow-tie to its current position, gives a modestly higher value of $\sim 10~\mathrm{kyrs}$ (see Sect. \ref{sec:association}). On the other hand, the minimum energy stored in the relativistic plasma requires a total mass of $1.3 \times 10^{25}~\mathrm{g}$. The required energy corresponds to $\sim 8~\mathrm{yrs}$' accumulation at the Eddington rate for a $6.4~\mathrm{M}_\sun$ black hole, while the mass could be accumulated in $\sim 160~\mathrm{days}$. V4641~Sgr is active every 1--3 years, but its outbursts are much shorter than typical BH LMXBs \citep[e.g.,][]{MaitraBailyn2006}, making its duty cycle rather low. Taking an average value of 2 years for the recurrence time and 10 days for the outburst mean duration, the energy accumulation time converts to $\sim 600~\mathrm{yrs}$; namely, this is roughly five times smaller than the synchrotron cooling time and almost equal to the expansion time at $0.1c$. We thus conclude that the estimated age of the structure is compatible with the energy accumulation time, demonstrating the plausibility of the jet scenario in terms of timescales.

In the wind scenario, we can assume a similar energy budget, although in that case the expansion time will be much higher. Even with a wind speed of $3000~\mathrm{km~s^{-1}}$ \citep{Munoz-Darias2018}, the expansion time is as high as $\sim5.9~\mathrm{kyrs}$ (i.e., $\sim$twice the synchrotron cooling time), making the wind scenario less probable than the jet scenario in the synchrotron framework.

\subsubsection{Free-free}
\label{sec:freefree}

The bow-tie is reminiscent of the smaller scale "ruff" of equatorial emission that surrounds SS 433 \citep{Paragi1999, Paragi2002evn, Blundell2001}, as well as the parsec-scale faint diffuse structure recently observed around Cir~X--1 \citep{Cowie2025}. In the case of SS 433, the ruff was interpreted as thermal bremsstrahlung emission from a wind-like outflow, either from the companion, the accretion disk \citep{Blundell2002}, or a circumbinary disk \citep{Blundell2008, DoolinBlundell2009}. Similarly, the large-scale radio structure around V4641~Sgr could arise from free-free emission produced by a disk wind. V4641~Sgr has exhibited at least one super-Eddington outburst \citep{Revnivtsev2002}, implying significant radiatively driven mass-loss, likely in the form of equatorial disk winds. The free-free hypothesis implies to break the association between the radio and X-rays.

The bremsstrahlung spectral emissivity, $\varepsilon_\nu$, is related to the electron density, $n_e$, and temperature, $T$. Hence, we can use the flux density measurement and a range of temperatures to estimate whether reasonable ambient conditions can reproduce the observed emission. The electron density is expressed as\begin{equation*}
    n_e = \left[ \frac{\varepsilon_\nu T^{1/2} \exp(h\nu/k_B T)}{\kappa g(\nu, T)} \right]^{1/2},
\end{equation*}
where $\kappa \simeq 6.84\times 10^{-38}~\mathrm{erg~s^{-1}~Hz^{-1}~cm^3~K^{1/2}}$ and $g(\nu, T)$ is the Gaunt factor (see Appendix \ref{appendix:free-free} for a more detailed treatment). An average emissivity can be derived using the measured spectral luminosity, $L_\nu$, and a emission volume, $V$, as $\varepsilon_\nu = L_\nu / V = 4\pi D^2 S_\nu / V$.
We consider a temperature range of $10^4$--$10^7~\mathrm{K}$: below the lower bound, the gas is not fully ionized and, thus, cannot radiate efficiently through bremsstrahlung. These temperatures give Gaunt factors in the range 6--12 and constrain the electron density between 2 and 11 $\mathrm{cm}^{-3}$ (modulated by a factor of $(\sin\theta)^{1/2}$), which are typical values in the ISM. We therefore conclude that bremsstrahlung radiation is able to produce the observed radio luminosity under reasonable ambient conditions.

In the context of radio-quiet quasars, \citet{BlundellKuncic2007} developed a model for the core radio emission, in which thermal disk winds contribute to the flat radio spectrum through optically thin free-free radiation. This model predicts the specific radio luminosity as a function of the mass outflow rate, $\dot{m}_w$, wind velocity, $v_w$, and electron temperature, $T$, expressed as
\begin{equation*}
    L_\nu = \kappa g(\nu, T) (4\pi f_\Omega)^{-1} T^{-1/2} \left( \frac{\dot{m}_w}{\mu m_p v_w} \right)^{2} r_\mathrm{ph}^{-1},
\end{equation*}
where $\mu$ is the mean molecular weight of the plasma, $m_p$ is the mass of the proton, $f_\Omega = \Omega / 4\pi$ is the geometrical covering factor of the outflow\footnote{$\Omega$ is the solid angle of the outflow as seen from the central source.} and $r_\mathrm{ph}$ is the photospheric radius, namely, the radius beyond which the wind becomes transparent \citep{BlundellKuncic2007}. From both theory and observation, thermal disk winds in BH LMXBs likely have mass loss rates similar to the accretion rate \citep[e.g.,][]{Woods1996, Ponti2012, FenderMunoz-Darias2016, Higginbottom2019}. Hence, at Eddington rate, we have $\dot{m}_w \simeq \dot{m}_\mathrm{Edd} = 9 \times 10^{17}~\mathrm{g~s^{-1}}$. At this mass rate, the relevant expression for the photospheric radius is the free-free absorption radius,
\begin{equation*}
    r_\mathrm{ph} = r_\mathrm{ff} = A g(\nu, T)^{1/3} \nu^{-2/3} T^{-1/2} \left( \frac{\dot{m}_w}{4\pi f_\Omega \mu m_p v_w} \right)^{2/3},
\end{equation*}
where $A = (\kappa c^2 / 24 \pi k_B)^{1/3} \simeq 0.18  ~ \mathrm{cm^{5/3} Hz^{2/3} K^{1/2}}$. Using the expression for $r_\mathrm{ph}$, the spectral luminosity becomes
\begin{equation*}
    L_\nu = \frac{\kappa}{A} g(\nu, T)^{2/3} (4\pi f_\Omega)^{-1/3} \nu^{2/3} \left( \frac{\dot{m}_w}{\mu m_p v_w} \right)^{4/3}.
\end{equation*}
Previous studies have measured wind terminal velocities in the range 900--1600 $\mathrm{km~s^{-1}}$ \citep[e.g.,][]{Munoz-Darias2018}; thus, we chose $v_w = 1000~\mathrm{km~s^{-1}}$. Finally, we estimated $f_\Omega = 0.3$ from the opening angle of the bow-tie. At $\nu = 1.2~\mathrm{GHz}$ and assuming a fully ionized hydrogen plasma ($\mu = 0.5$), the resulting spectral luminosity is $L_\nu = (2.3-3.4) \times 10^{15}~\mathrm{erg~s^{-1}~Hz^{-1}}$ for the same temperature range as above. On the other hand, we were able to extract the luminosity of the bow-tie from our observations and found $L_\nu = 4\pi D^2 S_\nu \simeq 4.0 \times 10^{21}~\mathrm{erg~s^{-1}~Hz^{-1}}$. This measurement is six orders of magnitude higher than the theoretical value, indicating that the model does not successfully explain the radio luminosity. Even with super-Eddington accretion (e.g., $\dot{m} = 10~\dot{m}_\mathrm{Edd}$) and disk winds with velocities reduced by a factor of 10, the expected luminosity is still off by a factor of 2000. This result argues against a "pure wind" origin, namely, a continually replenished wind plasma emitting via optically thin free-free emission.

An estimate of the total energy input by a potential jet or wind can be computed using the enthalpy of the region, since it encapsulates the internal energy of the gas and the work done in blowing the bow-tie. For an ideal non-relativistic gas, the enthalpy is given by $H = 5PV/2$, while the pressure can be expressed as $P = n_ek_BT$. For the same range in temperature ($10^4$--$10^7$ K), we find the pressure to be $P = 3.2 \times 10^{-12} - 1.3 \times 10^{-8}~\mathrm{erg~cm^{-3}}$ and the enthalpy $H = 1.2 \times 10^{48} - 4.9\times 10^{51}~\mathrm{erg}$ (modulated by a factor of $(\sin\theta)^{-1}$). At the Eddington rate, these energies could be accumulated in $\sim$0.05--190 kyrs, which converts into $\sim$3.4 kyrs--14 Myrs when considering the same duty cycle as in Sect. \ref{sec:synchrotron} (10 days of activity every 2 years). These energies (and timescales) are between one and three orders of magnitude higher than those derived when considering synchrotron, further decreasing the likelihood that free-free is the dominant emission mechanism.

\section{Conclusions}
\label{sec:conclusions}

We  report the discovery of a faint, symmetric, large-scale ($\sim 35~\mathrm{pc}$) radio structure around the black hole X-ray binary V4641~Sgr, revealed through deep MeerKAT observations. This extended emission exhibits a bow-tie morphology and spatially coincides with X-ray emission previously detected by XRISM. While synchrotron emission from relativistic electrons accelerated in past outflow activity offers the most self-consistent explanation for the observed radio and X-ray properties, we also explored a free-free emission scenario associated with a thermal disk wind. Although this mechanism is able to reproduce the radio flux under plausible physical conditions, it does not account for the X-ray emission and requires significantly higher energy input. The geometry and energetics favor an origin linked to large-scale jets from V4641~Sgr, although a contribution from equatorial disk winds cannot be fully excluded. No clear radio counterpart is found for the larger-scale TeV gamma-ray structure detected by HAWC and LHAASO, indicating distinct emission regions and possibly different particle populations.

This detection adds V4641~Sgr to the growing population of black hole LMXBs exhibiting parsec-scale jets and extended emission, reinforcing the idea that microquasars can significantly impact their surrounding environment. Together with recent detections of VHE gamma-ray emission, these findings position V4641~Sgr as a promising laboratory to study particle acceleration and feedback processes from stellar-mass black holes.

\section*{Data availability}

The MeerKAT map shown in Figure \ref{fig:V4641_map} is only available in electronic form at the CDS via anonymous ftp to \url{cdsarc.u-strasbg.fr} (130.79.128.5) or via \url{http://cdsweb.u-strasbg.fr/cgi-bin/qcat?J/A+A/}.

\begin{acknowledgements}
We thank the referee for their careful reading and very fruitful comments, in particular regarding the proper motion study. The MeerKAT telescope is operated by the South African Radio Astronomy Observatory, which is a facility of the National Research Foundation, an agency of the Department of Science, Technology and Innovation. We acknowledge the use of the ilifu cloud computing facility - \url{www.ilifu.ac.za}, a partnership between the University of Cape Town, the University of the Western Cape, Stellenbosch University, Sol Plaatje University, the Cape Peninsula University of Technology and the South African Radio Astronomy Observatory. The ilifu facility is supported by contributions from the Inter-University Institute for Data Intensive Astronomy (IDIA - a partnership between the University of Cape Town, the University of Pretoria and the University of the Western Cape), the Computational Biology division at UCT and the Data Intensive Research Initiative of South Africa (DIRISA). This work made use of the CARTA (Cube Analysis and Rendering Tool for Astronomy) software (DOI 10.5281/zenodo.3377984 – \url{https://cartavis.github.io}). JHM acknowledges funding from a Royal Society University Research Fellowship (URF\textbackslash R1\textbackslash221062). The authors would like to thank Lilia Tremou, Andrew Hughes, Francesco Carotenuto and Payaswini Saikia for scheduling the MeerKAT observations and Fraser Cowie for providing access to the archival ThunderKAT data. Finally, we thank Pierre Kervella for his kind assistance with the Gaia astrometric data.
\end{acknowledgements}

\bibliographystyle{aa}
\bibliography{bib}

@article{Fender2018,
  author = "Fender, Rob  and Woudt, Patrick Alan  and Corbel, Stephane  and  Coriat, Mickaël  and  Daigne, Frédéric  and  Falcke, Heino  and  Girard, Julien  and  Heywood, Ian  and  Horesh, Assaf  and  Horrell, Jasper  and  Jonker, Peter G.  and  Joseph, Tana  and  Kamble, Atish  and  Knigge, Christian  and  Körding, Elmar  and  Kotze, Marissa  and  Kouveliotou, Chryssa  and  Lynch, Christine  and  Maccarone, Tom  and  Meintjes, Pieter  and  Migliari, Simone  and  Murphy, Tara  and  Nagayama, Takahiro  and  Nelemans, Gijs  and  Nicholson, George  and  O’Brien, Tim  and  Oodendaal, Alida  and  Oozeer, Nadeem  and  Osborne, Julian  and  Perez-Torres, Miguel  and  Ratcliffe, Simon  and  Ribeiro, Valério A.R.M.  and  Rol, Evert  and  Rushton, Anthony  and  Scaife, Anna  and  Schurch, Matthew  and  Sivakoff, Greg  and  Staley, Tim  and  Steeghs, Danny  and  Stewart, Ian  and  Swinbank, John D.  and  Vergani, Susanna  and  Warner, Brian  and  Wiersema, Klaas  and  Armstrong, Richard  and  Groot, Paul  and  McBride, Vanessa  and  Miller-Jones, James C.A.  and  Mooley, Kunal  and  Stappers, Ben  and  Wijers, Ralph A.M.J.  and  Bietenholz, Michael  and  Blyth, Sarah  and  Böttcher, Markus  and  Buckley, David  and  Charles, Phil  and  Chomiuk, Laura  and  Coppejans, Deanne  and  de Blok, W.J.G.  and  van der Heyden, Kurt  and  van der Horst, Alexander  and  van Soelen, Brian",
  title = "{ThunderKAT: The MeerKAT Large Survey Project for Image-Plane Radio Transients}",
  doi = "10.22323/1.277.0013",
  journal = "PoS",
  year = 2018,
  volume = "MeerKAT2016",
  pages = "013"
}

@software{oxkat,
       author = {{Heywood}, Ian},
        title = "{oxkat: Semi-automated imaging of MeerKAT observations}",
 howpublished = {Astrophysics Source Code Library, record ascl:2009.003},
         year = 2020,
        month = sep,
          eid = {ascl:2009.003},
       adsurl = {https://ui.adsabs.harvard.edu/abs/2020ascl.soft09003H}
}

@INPROCEEDINGS{casa,
       author = {{McMullin}, J.~P. and {Waters}, B. and {Schiebel}, D. and {Young}, W. and {Golap}, K.},
        title = "{CASA Architecture and Applications}",
    booktitle = {Astronomical Data Analysis Software and Systems XVI},
         year = 2007,
       editor = {{Shaw}, R.~A. and {Hill}, F. and {Bell}, D.~J.},
       series = {ASP Conf. Ser.},
       volume = {376},
        month = oct,
        pages = {127}
}

@INPROCEEDINGS{tricolour,
       author = {{Hugo}, Benjamin V. and {Perkins}, S. and {Merry}, B. and {Mauch}, T. and {Smirnov}, O.~M.},
        title = "{Tricolour: An Optimized SumThreshold Flagger for MeerKAT}",
    booktitle = {ASP},
         year = 2022,
       editor = {{Ruiz}, Jose Enrique and {Pierfedereci}, Francesco and {Teuben}, Peter},
       series = {ASP Conference Series},
       volume = {532},
        month = jul,
        pages = {541}
}

@ARTICLE{wsclean,
       author = {{Offringa}, A.~R. and {McKinley}, B. and {Hurley-Walker}, N. and {Briggs}, F.~H. and {Wayth}, R.~B. and {Kaplan}, D.~L. and {Bell}, M.~E. and {Feng}, L. and {Neben}, A.~R. and {Hughes}, J.~D. and {Rhee}, J. and {Murphy}, T. and {Bhat}, N.~D.~R. and {Bernardi}, G. and {Bowman}, J.~D. and {Cappallo}, R.~J. and {Corey}, B.~E. and {Deshpande}, A.~A. and {Emrich}, D. and {Ewall-Wice}, A. and {Gaensler}, B.~M. and {Goeke}, R. and {Greenhill}, L.~J. and {Hazelton}, B.~J. and {Hindson}, L. and {Johnston-Hollitt}, M. and {Jacobs}, D.~C. and {Kasper}, J.~C. and {Kratzenberg}, E. and {Lenc}, E. and {Lonsdale}, C.~J. and {Lynch}, M.~J. and {McWhirter}, S.~R. and {Mitchell}, D.~A. and {Morales}, M.~F. and {Morgan}, E. and {Kudryavtseva}, N. and {Oberoi}, D. and {Ord}, S.~M. and {Pindor}, B. and {Procopio}, P. and {Prabu}, T. and {Riding}, J. and {Roshi}, D.~A. and {Shankar}, N. Udaya and {Srivani}, K.~S. and {Subrahmanyan}, R. and {Tingay}, S.~J. and {Waterson}, M. and {Webster}, R.~L. and {Whitney}, A.~R. and {Williams}, A. and {Williams}, C.~L.},
        title = "{WSCLEAN: an implementation of a fast, generic wide-field imager for radio astronomy}",
      journal = {\mnras},
         year = 2014,
        month = oct,
       volume = {444},
       number = {1},
        pages = {606-619},
          doi = {10.1093/mnras/stu1368},
archivePrefix = {arXiv},
       eprint = {1407.1943},
 primaryClass = {astro-ph.IM},
       adsurl = {https://ui.adsabs.harvard.edu/abs/2014MNRAS.444..606O}
}

@ARTICLE{cubical,
       author = {{Kenyon}, J.~S. and {Smirnov}, O.~M. and {Grobler}, T.~L. and {Perkins}, S.~J.},
        title = "{CUBICAL - fast radio interferometric calibration suite exploiting complex optimization}",
      journal = {\mnras},
         year = 2018,
        month = aug,
       volume = {478},
       number = {2},
        pages = {2399-2415},
          doi = {10.1093/mnras/sty1221},
archivePrefix = {arXiv},
       eprint = {1805.03410},
 primaryClass = {astro-ph.IM},
       adsurl = {https://ui.adsabs.harvard.edu/abs/2018MNRAS.478.2399K}
}

@article{Remillard_McClintock2006,
   author = "Remillard, Ronald A. and McClintock, Jeffrey E.",
   title = "X-Ray Properties of Black-Hole Binaries", 
   journal= "\araa",
   year = "2006",
   volume = "44",
   number = "Volume 44, 2006",
   pages = "49-92",
   doi = "https://doi.org/10.1146/annurev.astro.44.051905.092532",
   url = "https://www.annualreviews.org/content/journals/10.1146/annurev.astro.44.051905.092532",
   publisher = "Annual Reviews",
   issn = "1545-4282",
   type = "Journal Article",
  }

@Inbook{Belloni_Motta2016,
author="Belloni, Tomaso M.
and Motta, Sara E.",
title="Transient Black Hole Binaries",
bookTitle="Astrophysics of Black Holes: From Fundamental Aspects to Latest Developments",
year="2016",
publisher="Springer Berlin Heidelberg",
address="Berlin, Heidelberg",
pages="61--97",
isbn="978-3-662-52859-4"
}

@article{Mirabel1994,
  title = {A superluminal source in the Galaxy},
  volume = {371},
  ISSN = {1476-4687},
  url = {http://dx.doi.org/10.1038/371046a0},
  DOI = {10.1038/371046a0},
  number = {6492},
  journal = {Nature},
  publisher = {Springer Science and Business Media LLC},
  author = {Mirabel,  I. F. and Rodríguez,  L. F.},
  year = {1994},
  month = sep,
  pages = {46–48}
}

@article{Russell2019,
  title = {Disk–Jet Coupling in the 2017/2018 Outburst of the Galactic Black Hole Candidate X-Ray Binary MAXI J1535–571},
  volume = {883},
  ISSN = {1538-4357},
  url = {http://dx.doi.org/10.3847/1538-4357/ab3d36},
  DOI = {10.3847/1538-4357/ab3d36},
  number = {2},
  journal = {\apj},
  publisher = {American Astronomical Society},
  author = {Russell,  T. D. and Tetarenko,  A. J. and Miller-Jones,  J. C. A. and Sivakoff,  G. R. and Parikh,  A. S. and Rapisarda,  S. and Wijnands,  R. and Corbel,  S. and Tremou,  E. and Altamirano,  D. and Baglio,  M. C. and Ceccobello,  C. and Degenaar,  N. and Eijnden,  J. van den and Fender,  R. and Heywood,  I. and Krimm,  H. A. and Lucchini,  M. and Markoff,  S. and Russell,  D. M. and Soria,  R. and Woudt,  P. A.},
  year = {2019},
  month = oct,
  pages = {198}
}

@article{Hjellming1995,
  title = {Episodic ejection of relativistic jets by the X-ray transient GRO J1655 - 40},
  volume = {375},
  ISSN = {1476-4687},
  url = {http://dx.doi.org/10.1038/375464a0},
  DOI = {10.1038/375464a0},
  number = {6531},
  journal = {Nature},
  publisher = {Springer Science and Business Media LLC},
  author = {Hjellming,  R. M. and Rupen,  M. P.},
  year = {1995},
  month = jun,
  pages = {464–468}
}

@article{Corbel2002,
author = {S. Corbel  and R. P. Fender  and A. K. Tzioumis  and J. A. Tomsick  and J. A. Orosz  and J. M. Miller  and R. Wijnands  and P. Kaaret },
title = {Large-Scale, Decelerating, Relativistic X-ray Jets from the Microquasar XTE J1550-564},
journal = {Science},
volume = {298},
number = {5591},
pages = {196-199},
year = {2002},
doi = {10.1126/science.1075857},
URL = {https://www.science.org/doi/abs/10.1126/science.1075857},
eprint = {https://www.science.org/doi/pdf/10.1126/science.1075857}}

@article{Bright2020,
  title = {An extremely powerful long-lived superluminal ejection from the black hole MAXI J1820+070},
  volume = {4},
  ISSN = {2397-3366},
  url = {http://dx.doi.org/10.1038/s41550-020-1023-5},
  DOI = {10.1038/s41550-020-1023-5},
  number = {7},
  journal = {Nature Astronomy},
  publisher = {Springer Science and Business Media LLC},
  author = {Bright,  J. S. and Fender,  R. P. and Motta,  S. E. and Williams,  D. R. A. and Moldon,  J. and Plotkin,  R. M. and Miller-Jones,  J. C. A. and Heywood,  I. and Tremou,  E. and Beswick,  R. and Sivakoff,  G. R. and Corbel,  S. and Buckley,  D. A. H. and Homan,  J. and Gallo,  E. and Tetarenko,  A. J. and Russell,  T. D. and Green,  D. A. and Titterington,  D. and Woudt,  P. A. and Armstrong,  R. P. and Groot,  P. J. and Horesh,  A. and Horst,  A. J. van der and K\"{o}rding,  E. G. and McBride,  V. A. and Rowlinson,  A. and Wijers,  R. A. M. J.},
  year = {2020},
  month = mar,
  pages = {697–703}
}

@article{Miller-Jones2012,
    author = {Miller-Jones, J. C. A. and Sivakoff, G. R. and Altamirano, D. and Coriat, M. and Corbel, S. and Dhawan, V. and Krimm, H. A. and Remillard, R. A. and Rupen, M. P. and Russell, D. M. and Fender, R. P. and Heinz, S. and Körding, E. G. and Maitra, D. and Markoff, S. and Migliari, S. and Sarazin, C. L. and Tudose, V.},
    title = "{Disc–jet coupling in the 2009 outburst of the black hole candidate H1743−322}",
    journal = {\mnras},
    volume = {421},
    number = {1},
    pages = {468-485},
    year = {2012},
    month = {03},
    issn = {0035-8711},
    doi = {10.1111/j.1365-2966.2011.20326.x},
    url = {https://doi.org/10.1111/j.1365-2966.2011.20326.x},
    eprint = {https://academic.oup.com/mnras/article-pdf/421/1/468/3134926/mnras0421-0468.pdf},
}

@ARTICLE{Alfaro2024,
       author = {{Alfaro}, R. and {Alvarez}, C. and {Arteaga-Vel{\'a}zquez}, J.~C. and {Avila Rojas}, D. and {Ayala Solares}, H.~A. and {Babu}, R. and {Belmont-Moreno}, E. and {Caballero-Mora}, K.~S. and {Capistr{\'a}n}, T. and {Carrami{\~n}ana}, A. and {Casanova}, S. and {Cotti}, U. and {Cotzomi}, J. and {Couti{\~n}o de Le{\'o}n}, S. and {De la Fuente}, E. and {Depaoli}, D. and {Di Lalla}, N. and {Diaz Hernandez}, R. and {Dingus}, B.~L. and {DuVernois}, M.~A. and {Durocher}, M. and {D{\'\i}az-V{\'e}lez}, J.~C. and {Engel}, K. and {Espinoza}, C. and {Fan}, K.~L. and {Fang}, K. and {Fraija}, N. and {Fraija}, S. and {Garc{\'\i}a-Gonz{\'a}lez}, J.~A. and {Garfias}, F. and {Gonzalez Mu{\~n}oz}, A. and {Gonz{\'a}lez}, M.~M. and {Goodman}, J.~A. and {Groetsch}, S. and {Harding}, J.~P. and {Herzog}, I. and {Hinton}, J. and {Huang}, D. and {Hueyotl-Zahuantitla}, F. and {H{\"u}ntemeyer}, P. and {Iriarte}, A. and {Joshi}, V. and {Kaufmann}, S. and {Kieda}, D. and {de Le{\'o}n}, C. and {Lee}, J. and {Le{\'o}n Vargas}, H. and {Linnemann}, J.~T. and {Longinotti}, A.~L. and {Luis-Raya}, G. and {Malone}, K. and {Martinez}, O. and {Mart{\'\i}nez-Castro}, J. and {Matthews}, J.~A. and {Miranda-Romagnoli}, P. and {Morales-Soto}, J.~A. and {Moreno}, E. and {Mostaf{\'a}}, M. and {Nayerhoda}, A. and {Nellen}, L. and {Newbold}, M. and {Nisa}, M.~U. and {Noriega-Papaqui}, R. and {Olivera-Nieto}, L. and {Omodei}, N. and {Osorio}, M. and {P{\'e}rez Araujo}, Y. and {P{\'e}rez-P{\'e}rez}, E.~G. and {Rho}, C.~D. and {Rosa-Gonz{\'a}lez}, D. and {Ruiz-Velasco}, E. and {Salazar}, H. and {Salazar-Gallegos}, D. and {Sandoval}, A. and {Schneider}, M. and {Serna-Franco}, J. and {Smith}, A.~J. and {Son}, Y. and {Springer}, R.~W. and {Tibolla}, O. and {Tollefson}, K. and {Torres}, I. and {Torres-Escobedo}, R. and {Turner}, R. and {Ure{\~n}a-Mena}, F. and {Varela}, E. and {Villase{\~n}or}, L. and {Wang}, X. and {Watson}, I.~J. and {Willox}, E. and {Yun-C{\'a}rcamo}, S. and {Zhou}, H.},
        title = "{Ultra-high-energy gamma-ray bubble around microquasar V4641 Sgr}",
      journal = {\nat},
         year = 2024,
        month = oct,
       volume = {634},
       number = {8034},
        pages = {557-560},
          doi = {10.1038/s41586-024-07995-9},
archivePrefix = {arXiv},
       eprint = {2410.16117},
 primaryClass = {astro-ph.HE},
       adsurl = {https://ui.adsabs.harvard.edu/abs/2024Natur.634..557A}
}

@article{LHAASO2025,
    author = {{LHAASO Collaboration} and Cao, Zhen and Aharonian, F and Bai, Y X and Bao, Y W and Bastieri, D and Bi, X J and Bi, Y J and Bian, W and Bukevich, A V and Cai, C M and Cao, W Y and Cao, Zhe and Chang, J and Chang, J F and Chen, A M and Chen, E S and Chen, G H and Chen, H X and Chen, Liang and Chen, Long and Chen, M J and Chen, M L and Chen, Q H and Chen, S and Chen, S H and Chen, S Z and Chen, T L and Chen, X B and Chen, X J and Chen, Y and Cheng, N and Cheng, Y D and Chu, M C and Cui, M Y and Cui, S W and Cui, X H and Cui, Y D and Dai, B Z and Dai, H L and Dai, Z G and Danzengluobu and Diao, Y X and Dong, X Q and Duan, K K and Fan, J H and Fan, Y Z and Fang, J and Fang, J H and Fang, K and Feng, C F and Feng, H and Feng, L and Feng, S H and Feng, X T and Feng, Y and Feng, Y L and Gabici, S and Gao, B and Gao, C D and Gao, Q and Gao, W and Gao, W K and Ge, M M and Ge, T T and Geng, L S and Giacinti, G and Gong, G H and Gou, Q B and Gu, M H and Guo, F L and Guo, J and Guo, X L and Guo, Y Q and Guo, Y Y and Han, Y A and Hannuksela, O A and Hasan, M and He, H H and He, H N and He, J Y and He, X Y and He, Y and Hernández-Cadena, S and Hou, B W and Hou, C and Hou, X and Hu, H B and Hu, S C and Huang, C and Huang, D H and Huang, J J and Huang, T Q and Huang, W J and Huang, X T and Huang, X Y and Huang, Y and Huang, Y Y and Ji, X L and Jia, H Y and Jia, K and Jiang, H B and Jiang, K and Jiang, X W and Jiang, Z J and Jin, M and Kaci, S and Kang, M M and Karpikov, I and Khangulyan, D and Kuleshov, D and Kurinov, K and Li, B B and Li, Cheng and Li, Cong and Li, D and Li, F and Li, H B and Li, H C and Li, Jian and Li, Jie and Li, K and Li, L and Li, R L and Li, S D and Li, T Y and Li, W L and Li, X R and Li, Xin and Li, Y and Li, Y Z and Li, Zhe and Li, Zhuo and Liang, E W and Liang, Y F and Lin, S J and Liu, B and Liu, C and Liu, D and Liu, D B and Liu, H and Liu, H D and Liu, J and Liu, J L and Liu, J R and Liu, M Y and Liu, R Y and Liu, S M and Liu, W and Liu, X and Liu, Y and Liu, Y and Liu, Y N and Lou, Y Q and Luo, Q and Luo, Y and Lv, H K and Ma, B Q and Ma, L L and Ma, X H and Mao, J R and Min, Z and Mitthumsiri, W and Mou, G B and Mu, H J and Neronov, A and Ng, K C Y and Ni, M Y and Nie, L and Ou, L J and Pattarakijwanich, P and Pei, Z Y and Qi, J C and Qi, M Y and Qin, J J and Raza, A and Ren, C Y and Ruffolo, D and Sáiz, A and Semikoz, D and Shao, L and Shchegolev, O and Shen, Y Z and Sheng, X D and Shi, Z D and Shu, F W and Song, H C and Stenkin, Yu V and Stepanov, V and Su, Y and Sun, D X and Sun, H and Sun, Q N and Sun, X N and Sun, Z B and Tabasam, N H and Takata, J and Tam, P H T and Tan, H B and Tang, Q W and Tang, R and Tang, Z B and Tian, W W and Tong, C N and Wan, L H and Wang, C and Wang, G W and Wang, H G and Wang, J C and Wang, K and Wang, Kai and Wang, Kai and Wang, L P and Wang, L Y and Wang, L Y and Wang, R and Wang, W and Wang, X G and Wang, X J and Wang, X Y and Wang, Y and Wang, Y D and Wang, Z H and Wang, Z X and Wang, Zheng and Wei, D M and Wei, J J and Wei, Y J and Wen, T and Weng, S S and Wu, C Y and Wu, H R and Wu, Q W and Wu, S and Wu, X F and Wu, Y S and Xi, S Q and Xia, J and Xia, J J and Xiang, G M and Xiao, D X and Xiao, G and Xin, Y L and Xing, Y and Xiong, D R and Xiong, Z and Xu, D L and Xu, R F and Xu, R X and Xu, W L and Xue, L and Yan, D H and Yan, J Z and Yan, T and Yang, C W and Yang, C Y and Yang, F F and Yang, L L and Yang, M J and Yang, R Z and Yang, W X and Yang, Z H and Yao, Z G and Ye, X A and Yin, L Q and Yin, N and You, X H and You, Z Y and Yu, Y H and Yuan, Q and Yue, H and Zeng, H D and Zeng, T X and Zeng, W and Zeng, X T and Zha, M and Zhang, B B and Zhang, B T and Zhang, C and Zhang, F and Zhang, H and Zhang, H M and Zhang, H Y and Zhang, J L and Zhang, Li and Zhang, P F and Zhang, P P and Zhang, R and Zhang, S R and Zhang, S S and Zhang, W Y and Zhang, X and Zhang, X P and Zhang, Yi and Zhang, Yong and Zhang, Z P and Zhao, J and Zhao, L and Zhao, L Z and Zhao, S P and Zhao, X H and Zhao, Z H and Zheng, F and Zhong, W J and Zhou, B and Zhou, H and Zhou, J N and Zhou, M and Zhou, P and Zhou, R and Zhou, X X and Zhou, X X and Zhu, B Y and Zhu, C G and Zhu, F R and Zhu, H and Zhu, K J and Zou, Y C and Zuo, X},
    title = {Ultrahigh-energy gamma-ray emission associated with black hole-jet systems},
    journal = {National Science Review},
    pages = {nwaf496},
    year = {2025},
    month = {11},
    issn = {2095-5138},
    doi = {10.1093/nsr/nwaf496},
    url = {https://doi.org/10.1093/nsr/nwaf496},
    eprint = {https://academic.oup.com/nsr/advance-article-pdf/doi/10.1093/nsr/nwaf496/65284677/nwaf496.pdf},
}

@ARTICLE{Neronov2025,
       author = {{Neronov}, A. and {Oikonomou}, F. and {Semikoz}, D.},
        title = "{Multimessenger signature of cosmic rays from the microquasar V4641 Sgr propagating along a Galactic magnetic field line}",
      journal = {\prd},
         year = 2025,
        month = may,
       volume = {111},
       number = {10},
          eid = {103025},
        pages = {103025},
          doi = {10.1103/PhysRevD.111.103025},
       adsurl = {https://ui.adsabs.harvard.edu/abs/2025PhRvD.111j3025N}
}

@ARTICLE{Suzuki2025,
       author = {{Suzuki}, Hiromasa and {Tsuji}, Naomi and {Kanemaru}, Yoshiaki and {Shidatsu}, Megumi and {Olivera-Nieto}, Laura and {Safi-Harb}, Samar and {Kimura}, Shigeo S. and {de la Fuente}, Eduardo and {Casanova}, Sabrina and {Mori}, Kaya and {Wang}, Xiaojie and {Kato}, Sei and {Tateishi}, Dai and {Uchiyama}, Hideki and {Tanaka}, Takaaki and {Uchida}, Hiroyuki and {Inoue}, Shun and {Huang}, Dezhi and {Lemoine-Goumard}, Marianne and {Miura}, Daiki and {Ogawa}, Shoji and {Kobayashi}, Shogo B. and {Done}, Chris and {Parra}, Maxime and {D{\'\i}az Trigo}, Maria and {Mu{\~n}oz-Darias}, Teo and {Armas Padilla}, Montserrat and {Tomaru}, Ryota and {Ueda}, Yoshihiro},
        title = "{Detection of Extended X-Ray Emission around the PeVatron Microquasar V4641 Sgr with XRISM}",
      journal = {\apjl},
         year = 2025,
        month = jan,
       volume = {978},
       number = {2},
          eid = {L20},
        pages = {L20},
          doi = {10.3847/2041-8213/ad9d11},
archivePrefix = {arXiv},
       eprint = {2412.08089},
 primaryClass = {astro-ph.HE},
       adsurl = {https://ui.adsabs.harvard.edu/abs/2025ApJ...978L..20S}
}

@ARTICLE{MacDonald2014,
       author = {{MacDonald}, Rachel K.~D. and {Bailyn}, Charles D. and {Buxton}, Michelle and {Cantrell}, Andrew G. and {Chatterjee}, Ritaban and {Kennedy-Shaffer}, Ross and {Orosz}, Jerome A. and {Markwardt}, Craig B. and {Swank}, Jean H.},
        title = "{The Black Hole Binary V4641 Sagitarii: Activity in Quiescence and Improved Mass Determinations}",
      journal = {\apj},
         year = 2014,
        month = mar,
       volume = {784},
       number = {1},
          eid = {2},
        pages = {2},
          doi = {10.1088/0004-637X/784/1/2},
archivePrefix = {arXiv},
       eprint = {1401.4190},
 primaryClass = {astro-ph.SR},
       adsurl = {https://ui.adsabs.harvard.edu/abs/2014ApJ...784....2M}
}

@ARTICLE{intZand1999,
       author = {{in 't Zand}, J. and {Heise}, J. and {Bazzano}, A. and {Cocchi}, M. and {di Ciolo}, L. and {Muller}, J.~M.},
        title = "{SAX J1819.3-2525}",
      journal = {\iaucirc},
         year = 1999,
        month = mar,
       volume = {7119},
        pages = {1},
       adsurl = {https://ui.adsabs.harvard.edu/abs/1999IAUC.7119....1I}
}

@ARTICLE{Markwardt1999,
       author = {{Markwardt}, C.~B. and {Swank}, J.~H. and {Marshall}, F.~E.},
        title = "{XTE J1819-254, XTE J1743-363, XTE J1710-281, XTE J1723-376}",
      journal = {\iaucirc},
         year = 1999,
        month = mar,
       volume = {7120},
        pages = {1},
       adsurl = {https://ui.adsabs.harvard.edu/abs/1999IAUC.7120....1M}
}

@ARTICLE{Gandhi2019,
       author = {{Gandhi}, Poshak and {Rao}, Anjali and {Johnson}, Michael A.~C. and {Paice}, John A. and {Maccarone}, Thomas J.},
        title = "{Gaia Data Release 2 distances and peculiar velocities for Galactic black hole transients}",
      journal = {\mnras},
         year = 2019,
        month = may,
       volume = {485},
       number = {2},
        pages = {2642-2655},
          doi = {10.1093/mnras/stz438},
archivePrefix = {arXiv},
       eprint = {1804.11349},
 primaryClass = {astro-ph.HE},
       adsurl = {https://ui.adsabs.harvard.edu/abs/2019MNRAS.485.2642G}
}

@ARTICLE{Hjellming2000,
       author = {{Hjellming}, R.~M. and {Rupen}, M.~P. and {Hunstead}, R.~W. and {Campbell-Wilson}, D. and {Mioduszewski}, A.~J. and {Gaensler}, B.~M. and {Smith}, D.~A. and {Sault}, R.~J. and {Fender}, R.~P. and {Spencer}, R.~E. and {de la Force}, C.~J. and {Richards}, A.~M.~S. and {Garrington}, S.~T. and {Trushkin}, S.~A. and {Ghigo}, F.~D. and {Waltman}, E.~B. and {McCollough}, M.},
        title = "{Light Curves and Radio Structure of the 1999 September Transient Event in V4641 Sagittarii (=XTE J1819-254=SAX J1819.3-2525)}",
      journal = {\apj},
         year = 2000,
        month = dec,
       volume = {544},
       number = {2},
        pages = {977-992},
          doi = {10.1086/317255},
       adsurl = {https://ui.adsabs.harvard.edu/abs/2000ApJ...544..977H}
}

@ARTICLE{Orosz2001,
       author = {{Orosz}, Jerome A. and {Kuulkers}, Erik and {van der Klis}, Michiel and {McClintock}, Jeffrey E. and {Garcia}, Michael R. and {Callanan}, Paul J. and {Bailyn}, Charles D. and {Jain}, Raj K. and {Remillard}, Ronald A.},
        title = "{A Black Hole in the Superluminal Source SAX J1819.3-2525 (V4641 Sgr)}",
      journal = {\apj},
         year = 2001,
        month = jul,
       volume = {555},
       number = {1},
        pages = {489-503},
          doi = {10.1086/321442},
archivePrefix = {arXiv},
       eprint = {astro-ph/0103045},
 primaryClass = {astro-ph},
       adsurl = {https://ui.adsabs.harvard.edu/abs/2001ApJ...555..489O}
}

@ARTICLE{Abeysekara2018,
       author = {{Abeysekara}, A.~U. and {Albert}, A. and {Alfaro}, R. and {Alvarez}, C. and {{\'A}lvarez}, J.~D. and {Arceo}, R. and {Arteaga-Vel{\'a}zquez}, J.~C. and {Avila Rojas}, D. and {Ayala Solares}, H.~A. and {Belmont-Moreno}, E. and {BenZvi}, S.~Y. and {Brisbois}, C. and {Caballero-Mora}, K.~S. and {Capistr{\'a}n}, T. and {Carrami{\~n}ana}, A. and {Casanova}, S. and {Castillo}, M. and {Cotti}, U. and {Cotzomi}, J. and {Couti{\~n}o de Le{\'o}n}, S. and {De Le{\'o}n}, C. and {De la Fuente}, E. and {D{\'\i}az-V{\'e}lez}, J.~C. and {Dichiara}, S. and {Dingus}, B.~L. and {DuVernois}, M.~A. and {Ellsworth}, R.~W. and {Engel}, K. and {Espinoza}, C. and {Fang}, K. and {Fleischhack}, H. and {Fraija}, N. and {Galv{\'a}n-G{\'a}mez}, A. and {Garc{\'\i}a-Gonz{\'a}lez}, J.~A. and {Garfias}, F. and {Gonz{\'a}lez-Mu{\~n}oz}, A. and {Gonz{\'a}lez}, M.~M. and {Goodman}, J.~A. and {Hampel-Arias}, Z. and {Harding}, J.~P. and {Hernandez}, S. and {Hinton}, J. and {Hona}, B. and {Hueyotl-Zahuantitla}, F. and {Hui}, C.~M. and {H{\"u}ntemeyer}, P. and {Iriarte}, A. and {Jardin-Blicq}, A. and {Joshi}, V. and {Kaufmann}, S. and {Kar}, P. and {Kunde}, G.~J. and {Lauer}, R.~J. and {Lee}, W.~H. and {Le{\'o}n Vargas}, H. and {Li}, H. and {Linnemann}, J.~T. and {Longinotti}, A.~L. and {Luis-Raya}, G. and {L{\'o}pez-Coto}, R. and {Malone}, K. and {Marinelli}, S.~S. and {Martinez}, O. and {Martinez-Castellanos}, I. and {Mart{\'\i}nez-Castro}, J. and {Matthews}, J.~A. and {Miranda-Romagnoli}, P. and {Moreno}, E. and {Mostaf{\'a}}, M. and {Nayerhoda}, A. and {Nellen}, L. and {Newbold}, M. and {Nisa}, M.~U. and {Noriega-Papaqui}, R. and {Pretz}, J. and {P{\'e}rez-P{\'e}rez}, E.~G. and {Ren}, Z. and {Rho}, C.~D. and {Rivi{\`e}re}, C. and {Rosa-Gonz{\'a}lez}, D. and {Rosenberg}, M. and {Ruiz-Velasco}, E. and {Salesa Greus}, F. and {Sandoval}, A. and {Schneider}, M. and {Schoorlemmer}, H. and {Seglar Arroyo}, M. and {Sinnis}, G. and {Smith}, A.~J. and {Springer}, R.~W. and {Surajbali}, P. and {Taboada}, I. and {Tibolla}, O. and {Tollefson}, K. and {Torres}, I. and {Vianello}, G. and {Villase{\~n}or}, L. and {Weisgarber}, T. and {Werner}, F. and {Westerhoff}, S. and {Wood}, J. and {Yapici}, T. and {Yodh}, G. and {Zepeda}, A. and {Zhang}, H. and {Zhou}, H.},
        title = "{Very-high-energy particle acceleration powered by the jets of the microquasar SS 433}",
      journal = {\nat},
         year = 2018,
        month = oct,
       volume = {562},
       number = {7725},
        pages = {82-85},
          doi = {10.1038/s41586-018-0565-5},
archivePrefix = {arXiv},
       eprint = {1810.01892},
 primaryClass = {astro-ph.HE},
       adsurl = {https://ui.adsabs.harvard.edu/abs/2018Natur.562...82A}
}

@ARTICLE{HESS2024,
       author = {{H.~E.~S.~S. Collaboration} and {Aharonian}, F. and {Ait Benkhali}, F. and {Aschersleben}, J. and {Ashkar}, H. and {Backes}, M. and {Barbosa Martins}, V. and {Batzofin}, R. and {Becherini}, Y. and {Berge}, D. and {Bernl{\"o}hr}, K. and {Bi}, B. and {B{\"o}ttcher}, M. and {Boisson}, C. and {Bolmont}, J. and {de Lavergne}, M. de Bony and {Borowska}, J. and {Bouyahiaoui}, M. and {Breuhaus}, M. and {Brose}, R. and {Brown}, A.~M. and {Brun}, F. and {Bruno}, B. and {Bulik}, T. and {Burger-Scheidlin}, C. and {Caroff}, S. and {Casanova}, S. and {Cecil}, R. and {Celic}, J. and {Cerruti}, M. and {Chand}, T. and {Chandra}, S. and {Chen}, A. and {Chibueze}, J. and {Chibueze}, O. and {Cotter}, G. and {Dai}, S. and {Mbarubucyeye}, J. Damascene and {Djannati-Ata{\"\i}}, A. and {Dmytriiev}, A. and {Doroshenko}, V. and {Egberts}, K. and {Einecke}, S. and {Ernenwein}, J. -P. and {Filipovic}, M. and {Fontaine}, G. and {F{\"u}{\ss}ling}, M. and {Funk}, S. and {Gabici}, S. and {Ghafourizadeh}, S. and {Giavitto}, G. and {Glawion}, D. and {Glicenstein}, J. -F. and {Grolleron}, G. and {Haerer}, L. and {Hinton}, J.~A. and {Hofmann}, W. and {Holch}, T.~L. and {Holler}, M. and {Horns}, D. and {Jamrozy}, M. and {Jankowsky}, F. and {Jardin-Blicq}, A. and {Joshi}, V. and {Jung-Richardt}, I. and {Kasai}, E. and {Katarzy{\'n}ski}, K. and {Khatoon}, R. and {Kh{\'e}lifi}, B. and {Klepser}, S. and {Klu{\'z}niak}, W. and {Komin}, Nu. and {Kosack}, K. and {Kostunin}, D. and {Kundu}, A. and {Lang}, R.~G. and {Le Stum}, S. and {Leitl}, F. and {Lemi{\`e}re}, A. and {Lenain}, J. -P. and {Leuschner}, F. and {Lohse}, T. and {Luashvili}, A. and {Lypova}, I. and {Mackey}, J. and {Malyshev}, D. and {Malyshev}, D. and {Marandon}, V. and {Marchegiani}, P. and {Marcowith}, A. and {Mart{\'\i}-Devesa}, G. and {Marx}, R. and {Mehta}, A. and {Mitchell}, A. and {Moderski}, R. and {Mohrmann}, L. and {Montanari}, A. and {Moulin}, E. and {Murach}, T. and {Nakashima}, K. and {de Naurois}, M. and {Niemiec}, J. and {Noel}, A. Priyana and {Ohm}, S. and {Olivera-Nieto}, L. and {de Ona Wilhelmi}, E. and {Ostrowski}, M. and {Panny}, S. and {Panter}, M. and {Parsons}, R.~D. and {Peron}, G. and {Prokhorov}, D.~A. and {P{\"u}hlhofer}, G. and {Punch}, M. and {Quirrenbach}, A. and {Reichherzer}, P. and {Reimer}, A. and {Reimer}, O. and {Ren}, H. and {Renaud}, M. and {Reville}, B. and {Rieger}, F. and {Rowell}, G. and {Rudak}, B. and {Ricarte}, H. Rueda and {Ruiz-Velasco}, E. and {Sahakian}, V. and {Salzmann}, H. and {Santangelo}, A. and {Sasaki}, M. and {Sch{\"a}fer}, J. and {Sch{\"u}ssler}, F. and {Schwanke}, U. and {Shapopi}, J.~N.~S. and {Sol}, H. and {Specovius}, A. and {Spencer}, S. and {Stawarz}, L. and {Steenkamp}, R. and {Steinmassl}, S. and {Steppa}, C. and {Streil}, K. and {Sushch}, I. and {Suzuki}, H. and {Takahashi}, T. and {Tanaka}, T. and {Taylor}, A.~M. and {Terrier}, R. and {Tsirou}, M. and {Tsuji}, N. and {Unbehaun}, T. and {van Eldik}, C. and {Vecchi}, M. and {Veh}, J. and {Venter}, C. and {Vink}, J. and {Wach}, T. and {Wagner}, S.~J. and {Werner}, F. and {White}, R. and {Wierzcholska}, A. and {Wong}, Yu Wun and {Zacharias}, M. and {Zargaryan}, D. and {Zdziarski}, A.~A. and {Zech}, A. and {Zouari}, S. and {{\.Z}ywucka}, N.},
        title = "{Acceleration and transport of relativistic electrons in the jets of the microquasar SS 433}",
      journal = {Science},
         year = 2024,
        month = jan,
       volume = {383},
       number = {6681},
        pages = {402-406},
          doi = {10.1126/science.adi2048},
archivePrefix = {arXiv},
       eprint = {2401.16019},
 primaryClass = {astro-ph.HE},
       adsurl = {https://ui.adsabs.harvard.edu/abs/2024Sci...383..402H}
}

@ARTICLE{Negoro2024ATel,
       author = {{Negoro}, H. and {Nakajima}, M. and {Fujiwara}, K. and {Kudo}, Y. and {Shibui}, H. and {Takagi}, K. and {Takahashi}, H. and {Tatano}, K. and {Nishio}, H. and {Mihara}, T. and {Kawamuro}, T. and {Yamada}, S. and {Wang}, S. and {Tamagawa}, T. and {Kawai}, N. and {Matsuoka}, M. and {Sakamoto}, T. and {Serino}, M. and {Sugita}, S. and {Kawakubo}, Y. and {Hiramatsu}, H. and {Nishikawa}, H. and {Kondo}, Y. and {Yoshida}, A. and {Tsuboi}, Y. and {Sugai}, H. and {Nagashima}, N. and {Shidatsu}, M. and {Niida}, Y. and {Takahashi}, I. and {Niwano}, M. and {Higuchi}, N. and {Yatsu}, Y. and {Nakahira}, S. and {Ueno}, S. and {Tomida}, H. and {Ishikawa}, M. and {Ogawa}, S. and {Kurihara}, M. and {Ueda}, Y. and {Okada}, Y. and {Yamauchi}, M. and {Otsuki}, Y. and {Hasegawa}, T. and {Nishio}, M. and {Yamaoka}, K. and {Sugizaki}, M. and {Iwakiri}, W.},
        title = "{MAXI/GSC detection of renewed activity of the black hole X-ray binary V4641 Sgr}",
      journal = {The Astronomer's Telegram},
         year = 2024,
        month = sep,
       volume = {16804},
        pages = {1},
       adsurl = {https://ui.adsabs.harvard.edu/abs/2024ATel16804....1N}
}

@ARTICLE{naima,
   author = {{Zabalza}, V.},
    title = {naima: a Python package for inference of relativistic particle
             energy distributions from observed nonthermal spectra},
     year = 2015,
  journal = {Proc.~of International Cosmic Ray Conference 2015},
    pages = "922",
   eprint = {1509.03319},
   adsurl = {http://adsabs.harvard.edu/abs/2015arXiv150903319Z},
}

@book{Longair2011, place={Cambridge}, edition={3}, title={High Energy Astrophysics}, publisher={Cambridge University Press}, author={Longair, Malcolm S.}, year={2011}}

@ARTICLE{Gallo2005,
       author = {{Gallo}, Elena and {Fender}, Rob and {Kaiser}, Christian and {Russell}, David and {Morganti}, Raffaella and {Oosterloo}, Tom and {Heinz}, Sebastian},
        title = "{A dark jet dominates the power output of the stellar black hole Cygnus X-1}",
      journal = {\nat},
         year = 2005,
        month = aug,
       volume = {436},
       number = {7052},
        pages = {819-821},
          doi = {10.1038/nature03879},
archivePrefix = {arXiv},
       eprint = {astro-ph/0508228},
 primaryClass = {astro-ph},
       adsurl = {https://ui.adsabs.harvard.edu/abs/2005Natur.436..819G}
}

@ARTICLE{Marti2017,
       author = {{Mart{\'\i}}, Josep and {Luque-Escamilla}, Pedro L. and {Bosch-Ramon}, Valent{\'\i} and {Paredes}, Josep M.},
        title = "{A galactic microquasar mimicking winged radio galaxies}",
      journal = {Nature Communications},
         year = 2017,
        month = nov,
       volume = {8},
          eid = {1757},
        pages = {1757},
          doi = {10.1038/s41467-017-01976-5},
       adsurl = {https://ui.adsabs.harvard.edu/abs/2017NatCo...8.1757M}
}

@ARTICLE{Mirabel1992,
       author = {{Mirabel}, I.~F. and {Rodriguez}, L.~F. and {Cordier}, B. and {Paul}, J. and {Lebrun}, F.},
        title = "{A double-sided radio jet from the compact Galactic Centre annihilator 1E1740.7-2942}",
      journal = {\nat},
         year = 1992,
        month = jul,
       volume = {358},
       number = {6383},
        pages = {215-217},
          doi = {10.1038/358215a0},
       adsurl = {https://ui.adsabs.harvard.edu/abs/1992Natur.358..215M}
}

@ARTICLE{Angelini2003,
       author = {{Angelini}, Lorella and {White}, Nicholas E.},
        title = "{An XMM-Newton Observation of 4U 1755-33 in Quiescence: Evidence of a Fossil X-Ray Jet}",
      journal = {\apjl},
         year = 2003,
        month = mar,
       volume = {586},
       number = {1},
        pages = {L71-L75},
          doi = {10.1086/374682},
archivePrefix = {arXiv},
       eprint = {astro-ph/0302315},
 primaryClass = {astro-ph},
       adsurl = {https://ui.adsabs.harvard.edu/abs/2003ApJ...586L..71A}
}

@ARTICLE{Kaaret2006,
       author = {{Kaaret}, P. and {Corbel}, S. and {Tomsick}, J.~A. and {Lazendic}, J. and {Tzioumis}, A.~K. and {Butt}, Y. and {Wijnands}, R.},
        title = "{Evolution of the X-Ray Jets from 4U 1755-33}",
      journal = {\apj},
         year = 2006,
        month = apr,
       volume = {641},
       number = {1},
        pages = {410-417},
          doi = {10.1086/500399},
archivePrefix = {arXiv},
       eprint = {astro-ph/0512240},
 primaryClass = {astro-ph},
       adsurl = {https://ui.adsabs.harvard.edu/abs/2006ApJ...641..410K}
}

@ARTICLE{MillerJones2006,
       author = {{Miller-Jones}, J.~C.~A. and {Fender}, R.~P. and {Nakar}, E.},
        title = "{Opening angles, Lorentz factors and confinement of X-ray binary jets}",
      journal = {\mnras},
         year = 2006,
        month = apr,
       volume = {367},
       number = {4},
        pages = {1432-1440},
          doi = {10.1111/j.1365-2966.2006.10092.x},
archivePrefix = {arXiv},
       eprint = {astro-ph/0601482},
 primaryClass = {astro-ph},
       adsurl = {https://ui.adsabs.harvard.edu/abs/2006MNRAS.367.1432M}
}

@ARTICLE{Gallo2014,
       author = {{Gallo}, Elena and {Plotkin}, Richard M. and {Jonker}, Peter G.},
        title = "{V4641 Sgr: a candidate precessing microblazar}",
      journal = {\mnras},
         year = 2014,
        month = feb,
       volume = {438},
       number = {1},
        pages = {L41-L45},
          doi = {10.1093/mnrasl/slt152},
archivePrefix = {arXiv},
       eprint = {1310.7032},
 primaryClass = {astro-ph.HE},
       adsurl = {https://ui.adsabs.harvard.edu/abs/2014MNRAS.438L..41G}
}

@ARTICLE{Salvesen2020,
       author = {{Salvesen}, Greg and {Pokawanvit}, Supavit},
        title = "{Origin of spin-orbit misalignments: the microblazar V4641 Sgr}",
      journal = {\mnras},
         year = 2020,
        month = jun,
       volume = {495},
       number = {2},
        pages = {2179-2204},
          doi = {10.1093/mnras/staa1094},
archivePrefix = {arXiv},
       eprint = {2004.08392},
 primaryClass = {astro-ph.HE},
       adsurl = {https://ui.adsabs.harvard.edu/abs/2020MNRAS.495.2179S}
}

@article{Munoz-Darias2018,
    author = {Muñoz-Darias, Teo and Torres, Manuel A P and Garcia, Michael R},
    title = {The low-luminosity accretion disc wind of the black hole transient V4641 Sagittarii},
    journal = {\mnras},
    volume = {479},
    number = {3},
    pages = {3987-3995},
    year = {2018},
    month = {06},
    issn = {0035-8711},
    doi = {10.1093/mnras/sty1711},
    url = {https://doi.org/10.1093/mnras/sty1711},
    eprint = {https://academic.oup.com/mnras/article-pdf/479/3/3987/25169892/sty1711.pdf},
}

@article{Chaty2003,
    author = {Chaty, S. and Charles, P. A. and Martí, J. and Mirabel, I. F. and Rodríguez, L. F. and Shahbaz, T.},
    title = {Optical and near-infrared observations of the microquasar V4641 Sgr during the 1999 September outburst},
    journal = {\mnras},
    volume = {343},
    number = {1},
    pages = {169-174},
    year = {2003},
    month = {07},
    issn = {0035-8711},
    doi = {10.1046/j.1365-8711.2003.06651.x},
    url = {https://doi.org/10.1046/j.1365-8711.2003.06651.x},
    eprint = {https://academic.oup.com/mnras/article-pdf/343/1/169/3443354/343-1-169.pdf},
}

@ARTICLE{Corbel2004,
       author = {{Corbel}, S. and {Fender}, R.~P. and {Tomsick}, J.~A. and {Tzioumis}, A.~K. and {Tingay}, S.},
        title = "{On the Origin of Radio Emission in the X-Ray States of XTE J1650-500 during the 2001-2002 Outburst}",
      journal = {\apj},
         year = 2004,
        month = dec,
       volume = {617},
       number = {2},
        pages = {1272-1283},
          doi = {10.1086/425650},
archivePrefix = {arXiv},
       eprint = {astro-ph/0409154},
 primaryClass = {astro-ph},
       adsurl = {https://ui.adsabs.harvard.edu/abs/2004ApJ...617.1272C}
}

@article{Fender2004,
    author = {Fender, R. P. and Belloni, T. M. and Gallo, E.},
    title = {Towards a unified model for black hole X-ray binary jets},
    journal = {\mnras},
    volume = {355},
    number = {4},
    pages = {1105-1118},
    year = {2004},
    month = {12},
    issn = {0035-8711},
    doi = {10.1111/j.1365-2966.2004.08384.x},
    url = {https://doi.org/10.1111/j.1365-2966.2004.08384.x},
    eprint = {https://academic.oup.com/mnras/article-pdf/355/4/1105/6271740/355-4-1105.pdf},
}

@ARTICLE{MirabelRodriguez1999,
       author = {{Mirabel}, I.~F. and {Rodr{\'\i}guez}, L.~F.},
        title = "{Sources of Relativistic Jets in the Galaxy}",
      journal = {\araa},
         year = 1999,
        month = jan,
       volume = {37},
        pages = {409-443},
          doi = {10.1146/annurev.astro.37.1.409},
archivePrefix = {arXiv},
       eprint = {astro-ph/9902062},
 primaryClass = {astro-ph},
       adsurl = {https://ui.adsabs.harvard.edu/abs/1999ARA&A..37..409M}
}

@ARTICLE{Espinasse2020,
       author = {{Espinasse}, Mathilde and {Corbel}, St{\'e}phane and {Kaaret}, Philip and {Tremou}, Evangelia and {Migliori}, Giulia and {Plotkin}, Richard M. and {Bright}, Joe and {Tomsick}, John and {Tzioumis}, Anastasios and {Fender}, Rob and {Orosz}, Jerome A. and {Gallo}, Elena and {Homan}, Jeroen and {Jonker}, Peter G. and {Miller-Jones}, James C.~A. and {Russell}, David M. and {Motta}, Sara},
        title = "{Relativistic X-Ray Jets from the Black Hole X-Ray Binary MAXI J1820+070}",
      journal = {\apjl},
         year = 2020,
        month = jun,
       volume = {895},
       number = {2},
          eid = {L31},
        pages = {L31},
          doi = {10.3847/2041-8213/ab88b6},
archivePrefix = {arXiv},
       eprint = {2004.06416},
 primaryClass = {astro-ph.HE},
       adsurl = {https://ui.adsabs.harvard.edu/abs/2020ApJ...895L..31E}
}

@ARTICLE{Corbel2005,
       author = {{Corbel}, S. and {Kaaret}, P. and {Fender}, R.~P. and {Tzioumis}, A.~K. and {Tomsick}, J.~A. and {Orosz}, J.~A.},
        title = "{Discovery of X-Ray Jets in the Microquasar H1743-322}",
      journal = {\apj},
         year = 2005,
        month = oct,
       volume = {632},
       number = {1},
        pages = {504-513},
          doi = {10.1086/432499},
archivePrefix = {arXiv},
       eprint = {astro-ph/0505526},
 primaryClass = {astro-ph},
       adsurl = {https://ui.adsabs.harvard.edu/abs/2005ApJ...632..504C}
}

@ARTICLE{Kaaret2003,
       author = {{Kaaret}, P. and {Corbel}, S. and {Tomsick}, J.~A. and {Fender}, R. and {Miller}, J.~M. and {Orosz}, J.~A. and {Tzioumis}, A.~K. and {Wijnands}, R.},
        title = "{X-Ray Emission from the Jets of XTE J1550-564}",
      journal = {\apj},
         year = 2003,
        month = jan,
       volume = {582},
       number = {2},
        pages = {945-953},
          doi = {10.1086/344540},
archivePrefix = {arXiv},
       eprint = {astro-ph/0210401},
 primaryClass = {astro-ph},
       adsurl = {https://ui.adsabs.harvard.edu/abs/2003ApJ...582..945K}
}

@ARTICLE{Tomsick2003,
       author = {{Tomsick}, John A. and {Corbel}, St{\'e}phane and {Fender}, Rob and {Miller}, Jon M. and {Orosz}, Jerome A. and {Tzioumis}, Tasso and {Wijnands}, Rudy and {Kaaret}, Philip},
        title = "{X-Ray Jet Emission from the Black Hole X-Ray Binary XTE J1550-564 with Chandra in 2000}",
      journal = {\apj},
         year = 2003,
        month = jan,
       volume = {582},
       number = {2},
        pages = {933-944},
          doi = {10.1086/344703},
archivePrefix = {arXiv},
       eprint = {astro-ph/0210399},
 primaryClass = {astro-ph},
       adsurl = {https://ui.adsabs.harvard.edu/abs/2003ApJ...582..933T}
}

@ARTICLE{Migliori2017,
       author = {{Migliori}, Giulia and {Corbel}, S. and {Tomsick}, J.~A. and {Kaaret}, P. and {Fender}, R.~P. and {Tzioumis}, A.~K. and {Coriat}, M. and {Orosz}, J.~A.},
        title = "{Evolving morphology of the large-scale relativistic jets from XTE J1550-564}",
      journal = {\mnras},
         year = 2017,
        month = nov,
       volume = {472},
       number = {1},
        pages = {141-165},
          doi = {10.1093/mnras/stx1864},
archivePrefix = {arXiv},
       eprint = {1707.06876},
 primaryClass = {astro-ph.HE},
       adsurl = {https://ui.adsabs.harvard.edu/abs/2017MNRAS.472..141M}
}

@ARTICLE{Carotenuto2021a,
       author = {{Carotenuto}, F. and {Corbel}, S. and {Tremou}, E. and {Russell}, T.~D. and {Tzioumis}, A. and {Fender}, R.~P. and {Woudt}, P.~A. and {Motta}, S.~E. and {Miller-Jones}, J.~C.~A. and {Chauhan}, J. and {Tetarenko}, A.~J. and {Sivakoff}, G.~R. and {Heywood}, I. and {Horesh}, A. and {van der Horst}, A.~J. and {Koerding}, E. and {Mooley}, K.~P.},
        title = "{The black hole transient MAXI J1348-630: evolution of the compact and transient jets during its 2019/2020 outburst}",
      journal = {\mnras},
         year = 2021,
        month = jun,
       volume = {504},
       number = {1},
        pages = {444-468},
          doi = {10.1093/mnras/stab864},
archivePrefix = {arXiv},
       eprint = {2103.12190},
 primaryClass = {astro-ph.HE},
       adsurl = {https://ui.adsabs.harvard.edu/abs/2021MNRAS.504..444C}
}

@ARTICLE{Bahramian2023,
       author = {{Bahramian}, A. and {Tremou}, E. and {Tetarenko}, A.~J. and {Miller-Jones}, J.~C.~A. and {Fender}, R.~P. and {Corbel}, S. and {Williams}, D.~R.~A. and {Strader}, J. and {Carotenuto}, F. and {Salinas}, R. and {Kennea}, J.~A. and {Motta}, S.~E. and {Woudt}, P.~A. and {Matthews}, J.~H. and {Russell}, T.~D.},
        title = "{MAXI J1848-015: The First Detection of Relativistically Moving Outflows from a Globular Cluster X-Ray Binary}",
      journal = {\apjl},
         year = 2023,
        month = may,
       volume = {948},
       number = {1},
          eid = {L7},
        pages = {L7},
          doi = {10.3847/2041-8213/accde1},
archivePrefix = {arXiv},
       eprint = {2305.03764},
 primaryClass = {astro-ph.HE},
       adsurl = {https://ui.adsabs.harvard.edu/abs/2023ApJ...948L...7B}
}

@article{Nims2015,
    author = {Nims, Jesse and Quataert, Eliot and Faucher-Giguère, Claude-André},
    title = {Observational signatures of galactic winds powered by active galactic nuclei},
    journal = {\mnras},
    volume = {447},
    number = {4},
    pages = {3612-3622},
    year = {2015},
    month = {01},
    issn = {0035-8711},
    doi = {10.1093/mnras/stu2648},
    url = {https://doi.org/10.1093/mnras/stu2648},
    eprint = {https://academic.oup.com/mnras/article-pdf/447/4/3612/5696131/stu2648.pdf},
}

@ARTICLE{MaitraBailyn2006,
       author = {{Maitra}, Dipankar and {Bailyn}, Charles D.},
        title = "{X-Ray Observations of V4641 SGR (SAX J1819.3-2525) during the Brief and Violent Outburst of 2003}",
      journal = {\apj},
         year = 2006,
        month = feb,
       volume = {637},
       number = {2},
        pages = {992-1001},
          doi = {10.1086/498422},
archivePrefix = {arXiv},
       eprint = {astro-ph/0510531},
 primaryClass = {astro-ph},
       adsurl = {https://ui.adsabs.harvard.edu/abs/2006ApJ...637..992M}
}

@ARTICLE{HomanBelloni2005,
       author = {{Homan}, Jeroen and {Belloni}, Tomaso},
        title = "{The Evolution of Black Hole States}",
      journal = {\apss},
         year = 2005,
        month = nov,
       volume = {300},
       number = {1-3},
        pages = {107-117},
          doi = {10.1007/s10509-005-1197-4},
archivePrefix = {arXiv},
       eprint = {astro-ph/0412597},
 primaryClass = {astro-ph},
       adsurl = {https://ui.adsabs.harvard.edu/abs/2005Ap&SS.300..107H}
}

@INPROCEEDINGS{MacDonald2011,
       author = {{MacDonald}, Rachel K.~D. and {Bailyn}, C.~D. and {Cantrell}, A.~G.},
        title = "{Mass of the Black Hole in V4641 Sgr}",
    booktitle = {American Astronomical Society Meeting Abstracts \#217},
         year = 2011,
       series = {American Astronomical Society Meeting Abstracts},
       volume = {217},
        month = jan,
          eid = {144.20},
        pages = {144.20},
       adsurl = {https://ui.adsabs.harvard.edu/abs/2011AAS...21714420M}
}

@INPROCEEDINGS{Barsukova2014,
       author = {{Barsukova}, E. and {Goranskij}, V. and {Kroll}, P.},
        title = "{Historical light curve of the black hole binary V4641 Sgr based on the Moscow and Sonneberg plate archives}",
    booktitle = {Astroplate 2014},
         year = 2014,
        month = jan,
        pages = {99},
          doi = {10.48550/arXiv.1410.2055},
archivePrefix = {arXiv},
       eprint = {1410.2055},
 primaryClass = {astro-ph.HE},
       adsurl = {https://ui.adsabs.harvard.edu/abs/2014aspl.conf...99B}
}

@INPROCEEDINGS{Mioduszewski2003,
       author = {{Mioduszewski}, A.~J. and {Rupen}, M.~P. and {Walker}, R.~C. and {Taylor}, G.~B.},
        title = "{Unraveling a Precessing Jet: Forty Daily VLBA Observations of SS433}",
    booktitle = {American Astronomical Society Meeting Abstracts},
         year = 2003,
       series = {American Astronomical Society Meeting Abstracts},
       volume = {203},
        month = dec,
          eid = {31.05},
        pages = {31.05},
       adsurl = {https://ui.adsabs.harvard.edu/abs/2003AAS...203.3105M}
}

@ARTICLE{Motta2025,
       author = {{Motta}, S.~E. and {Atri}, P. and {Matthews}, James H. and {van den Eijnden}, Jakob and {Fender}, Rob P. and {Miller-Jones}, James C.~A. and {Heywood}, Ian and {Woudt}, Patrick},
        title = "{MeerKAT discovers a jet-driven bow shock near GRS 1915+105: How an invisible large-scale jet sculpts a microquasar's environment}",
      journal = {\aap},
         year = 2025,
        month = apr,
       volume = {696},
          eid = {A222},
        pages = {A222},
          doi = {10.1051/0004-6361/202452838},
archivePrefix = {arXiv},
       eprint = {2504.17425},
 primaryClass = {astro-ph.HE},
       adsurl = {https://ui.adsabs.harvard.edu/abs/2025A&A...696A.222M}
}

@ARTICLE{Atri2025,
       author = {{Atri}, P. and {Motta}, S.~E. and {van den Eijnden}, J. and {Matthews}, J.~H. and {Miller-Jones}, J.~C.~A. and {Fender}, R. and {Williams-Baldwin}, D. and {Heywood}, I. and {Woudt}, P.},
        title = "{Quantifying jet{\textendash}interstellar medium interactions in Cyg X-1: Insights from dual-frequency bow shock detection with MeerKAT}",
      journal = {\aap},
         year = 2025,
        month = apr,
       volume = {696},
          eid = {A223},
        pages = {A223},
          doi = {10.1051/0004-6361/202452837},
archivePrefix = {arXiv},
       eprint = {2504.17635},
 primaryClass = {astro-ph.HE},
       adsurl = {https://ui.adsabs.harvard.edu/abs/2025A&A...696A.223A}
}

@ARTICLE{Richards2021,
       author = {{Richards}, Gordon T. and {McCaffrey}, Trevor V. and {Kimball}, Amy and {Rankine}, Amy L. and {Matthews}, James H. and {Hewett}, Paul C. and {Rivera}, Angelica B.},
        title = "{Probing the Wind Component of Radio Emission in Luminous High-redshift Quasars}",
      journal = {\aj},
         year = 2021,
        month = dec,
       volume = {162},
       number = {6},
          eid = {270},
        pages = {270},
          doi = {10.3847/1538-3881/ac283b},
archivePrefix = {arXiv},
       eprint = {2106.07783},
 primaryClass = {astro-ph.GA},
       adsurl = {https://ui.adsabs.harvard.edu/abs/2021AJ....162..270R}
}

@ARTICLE{Rankine2021,
       author = {{Rankine}, Amy L. and {Matthews}, James H. and {Hewett}, Paul C. and {Banerji}, Manda and {Morabito}, Leah K. and {Richards}, Gordon T.},
        title = "{Placing LOFAR-detected quasars in C IV emission space: implications for winds, jets and star formation}",
      journal = {\mnras},
         year = 2021,
        month = apr,
       volume = {502},
       number = {3},
        pages = {4154-4169},
          doi = {10.1093/mnras/stab302},
archivePrefix = {arXiv},
       eprint = {2101.12635},
 primaryClass = {astro-ph.GA},
       adsurl = {https://ui.adsabs.harvard.edu/abs/2021MNRAS.502.4154R}
}

@ARTICLE{Stocke1992,
       author = {{Stocke}, John T. and {Morris}, Simon L. and {Weymann}, Ray J. and {Foltz}, Craig B.},
        title = "{The Radio Properties of the Broad-Absorption-Line QSOs}",
      journal = {\apj},
         year = 1992,
        month = sep,
       volume = {396},
        pages = {487},
          doi = {10.1086/171735},
       adsurl = {https://ui.adsabs.harvard.edu/abs/1992ApJ...396..487S}
}

@ARTICLE{Paragi1999,
       author = {{Paragi}, Z. and {Vermeulen}, R.~C. and {Fejes}, I. and {Schilizzi}, R.~T. and {Spencer}, R.~E. and {Stirling}, A.~M.},
        title = "{The inner radio jet region and the complex environment of SS433}",
      journal = {\aap},
         year = 1999,
        month = aug,
       volume = {348},
        pages = {910-916},
          doi = {10.48550/arXiv.astro-ph/9907169},
archivePrefix = {arXiv},
       eprint = {astro-ph/9907169},
 primaryClass = {astro-ph},
       adsurl = {https://ui.adsabs.harvard.edu/abs/1999A&A...348..910P}
}

@INPROCEEDINGS{Paragi2002evn,
       author = {{Paragi}, Z. and {Fejes}, I. and {Vermeulen}, R.~C. and {Schilizzi}, R.~T. and {Spencer}, R.~E. and {Stirling}, A.~M.},
        title = "{The Equatorial Outflow of SS 433}",
    booktitle = {Proceedings of the 6th EVN Symposium},
         year = 2002,
       editor = {{Ros}, Eduardo and {Porcas}, Richard W. and {Lobanov}, Andrei P. and {Zensus}, J. Anton},
        month = jun,
        pages = {263},
          doi = {10.48550/arXiv.astro-ph/0207061},
archivePrefix = {arXiv},
       eprint = {astro-ph/0207061},
 primaryClass = {astro-ph},
       adsurl = {https://ui.adsabs.harvard.edu/abs/2002evn..conf..263P}
}

@ARTICLE{Blundell2001,
       author = {{Blundell}, Katherine M. and {Mioduszewski}, Amy J. and {Muxlow}, Tom W.~B. and {Podsiadlowski}, Philipp and {Rupen}, Michael P.},
        title = "{Images of an Equatorial Outflow in SS 433}",
      journal = {\apjl},
         year = 2001,
        month = nov,
       volume = {562},
       number = {1},
        pages = {L79-L82},
          doi = {10.1086/324573},
archivePrefix = {arXiv},
       eprint = {astro-ph/0109504},
 primaryClass = {astro-ph},
       adsurl = {https://ui.adsabs.harvard.edu/abs/2001ApJ...562L..79B}
}

@inproceedings{Blundell2002,
    author = "Blundell, Katherine M and Rupen, Michael P and Mioduszewski, Amy J and Muxlow, Tom W. B. and Podsiadlowski, Philipp",
    title = "{The Ruff of equatorial emission around the SS433 jets: Its spectral index and origin}",
    booktitle = "{4th Microquasar Workshop: Microquasars and their Relation to Other Jet Sources in the Universe}",
    eprint = "astro-ph/0209365",
    archivePrefix = "arXiv",
    month = "9",
    year = "2002"
}

@ARTICLE{Blundell2008,
       author = {{Blundell}, Katherine M. and {Bowler}, Michael G. and {Schmidtobreick}, Linda},
        title = "{SS 433: Observation of the Circumbinary Disk and Extraction of the System Mass}",
      journal = {\apjl},
         year = 2008,
        month = may,
       volume = {678},
       number = {1},
        pages = {L47},
          doi = {10.1086/588027},
       adsurl = {https://ui.adsabs.harvard.edu/abs/2008ApJ...678L..47B}
}

@ARTICLE{DoolinBlundell2009,
       author = {{Doolin}, Samuel and {Blundell}, Katherine M.},
        title = "{The Precession of SS433's Radio Ruff on Long Timescales}",
      journal = {\apjl},
         year = 2009,
        month = jun,
       volume = {698},
       number = {1},
        pages = {L23-L26},
          doi = {10.1088/0004-637X/698/1/L23},
archivePrefix = {arXiv},
       eprint = {0905.1648},
 primaryClass = {astro-ph.SR},
       adsurl = {https://ui.adsabs.harvard.edu/abs/2009ApJ...698L..23D}
}

@ARTICLE{Revnivtsev2002,
       author = {{Revnivtsev}, M. and {Gilfanov}, M. and {Churazov}, E. and {Sunyaev}, R.},
        title = "{Super-Eddington outburst of V4641 Sgr}",
      journal = {\aap},
         year = 2002,
        month = sep,
       volume = {391},
        pages = {1013-1022},
          doi = {10.1051/0004-6361:20020865},
archivePrefix = {arXiv},
       eprint = {astro-ph/0204132},
 primaryClass = {astro-ph},
       adsurl = {https://ui.adsabs.harvard.edu/abs/2002A&A...391.1013R}
}

@ARTICLE{BlundellKuncic2007,
       author = {{Blundell}, Katherine M. and {Kuncic}, Zdenka},
        title = "{On the Origin of Radio Core Emission in Radio-quiet Quasars}",
      journal = {\apjl},
         year = 2007,
        month = oct,
       volume = {668},
       number = {2},
        pages = {L103-L106},
          doi = {10.1086/522695},
archivePrefix = {arXiv},
       eprint = {0708.2929},
 primaryClass = {astro-ph},
       adsurl = {https://ui.adsabs.harvard.edu/abs/2007ApJ...668L.103B}
}

@ARTICLE{Ponti2012,
       author = {{Ponti}, G. and {Fender}, R.~P. and {Begelman}, M.~C. and {Dunn}, R.~J.~H. and {Neilsen}, J. and {Coriat}, M.},
        title = "{Ubiquitous equatorial accretion disc winds in black hole soft states}",
      journal = {\mnras},
         year = 2012,
        month = may,
       volume = {422},
       number = {1},
        pages = {L11-L15},
          doi = {10.1111/j.1745-3933.2012.01224.x},
archivePrefix = {arXiv},
       eprint = {1201.4172},
 primaryClass = {astro-ph.HE},
       adsurl = {https://ui.adsabs.harvard.edu/abs/2012MNRAS.422L..11P}
}

@INCOLLECTION{FenderMunoz-Darias2016,
       author = {{Fender}, Rob and {Mu{\~n}oz-Darias}, Teo},
        title = "{The Balance of Power: Accretion and Feedback in Stellar Mass Black Holes}",
    booktitle = {Lecture Notes in Physics, Berlin Springer Verlag},
         year = 2016,
       editor = {{Haardt}, Francesco and {Gorini}, Vittorio and {Moschella}, Ugo and {Treves}, Aldo and {Colpi}, Monica},
       volume = {905},
        pages = {65},
          doi = {10.1007/978-3-319-19416-5_3},
       adsurl = {https://ui.adsabs.harvard.edu/abs/2016LNP...905...65F}
}

@ARTICLE{Woods1996,
       author = {{Woods}, D. Tod and {Klein}, Richard I. and {Castor}, John I. and {McKee}, Christopher F. and {Bell}, John B.},
        title = "{X-Ray--heated Coronae and Winds from Accretion Disks: Time-dependent Two-dimensional Hydrodynamics with Adaptive Mesh Refinement}",
      journal = {\apj},
         year = 1996,
        month = apr,
       volume = {461},
        pages = {767},
          doi = {10.1086/177101},
       adsurl = {https://ui.adsabs.harvard.edu/abs/1996ApJ...461..767W}
}

@ARTICLE{Higginbottom2019,
       author = {{Higginbottom}, Nick and {Knigge}, Christian and {Long}, Knox S. and {Matthews}, James H. and {Parkinson}, Edward J.},
        title = "{The luminosity dependence of thermally driven disc winds in low-mass X-ray binaries}",
      journal = {\mnras},
         year = 2019,
        month = apr,
       volume = {484},
       number = {4},
        pages = {4635-4644},
          doi = {10.1093/mnras/stz310},
archivePrefix = {arXiv},
       eprint = {1901.09684},
 primaryClass = {astro-ph.HE},
       adsurl = {https://ui.adsabs.harvard.edu/abs/2019MNRAS.484.4635H}
}

@ARTICLE{Wan2025arXiv,
       author = {{Wan}, Su-Yu and {Wang}, Jie-shuang and {Liu}, Ruo-Yu},
        title = "{A Leptonic Interpretation of the UHE Gamma-ray Emission from V4641 Sgr}",
      journal = {arXiv e-prints},
         year = 2025,
        month = jul,
          eid = {arXiv:2507.02763},
        pages = {arXiv:2507.02763},
          doi = {10.48550/arXiv.2507.02763},
archivePrefix = {arXiv},
       eprint = {2507.02763},
 primaryClass = {astro-ph.HE},
       adsurl = {https://ui.adsabs.harvard.edu/abs/2025arXiv250702763W}
}

@ARTICLE{Fender2023,
       author = {{Fender}, R.~P. and {Mooley}, K.~P. and {Motta}, S.~E. and {Bright}, J.~S. and {Williams}, D.~R.~A. and {Rushton}, A.~P. and {Beswick}, R.~J. and {Miller-Jones}, J.~C.~A. and {Kimura}, M. and {Isogai}, K. and {Kato}, T.},
        title = "{Comprehensive coverage of particle acceleration and kinetic feedback from the stellar mass black hole V404 Cygni}",
      journal = {\mnras},
         year = 2023,
        month = jan,
       volume = {518},
       number = {1},
        pages = {1243-1259},
          doi = {10.1093/mnras/stac1836},
archivePrefix = {arXiv},
       eprint = {2206.09831},
 primaryClass = {astro-ph.HE},
       adsurl = {https://ui.adsabs.harvard.edu/abs/2023MNRAS.518.1243F}
}

@ARTICLE{Cooper2025,
       author = {{Cooper}, A.~J. and {Matthews}, J.~H. and {Carotenuto}, F. and {Fender}, R. and {Lamb}, G.~P. and {Russell}, T.~D. and {Sarin}, N. and {Savard}, K. and {Zdziarski}, A.~A.},
        title = "{Joint Radiative and Kinematic Modelling of X-ray Binary Ejecta: Energy Estimate and Reverse Shock Detection}",
      journal = {\mnras},
         year = 2025,
        month = aug,
       volume = {541},
       number = {4},
        pages = {3518-3533},
          doi = {10.1093/mnras/staf1085},
archivePrefix = {arXiv},
       eprint = {2503.10804},
 primaryClass = {astro-ph.HE},
       adsurl = {https://ui.adsabs.harvard.edu/abs/2025MNRAS.541.3518C}
}

@ARTICLE{Matthews2025,
       author = {{Matthews}, James H. and {Cooper}, Alex J. and {Rhodes}, Lauren and {Savard}, Katherine and {Fender}, Rob and {Carotenuto}, Francesco and {Cowie}, Fraser J. and {Elley}, Emma L. and {Bright}, Joe and {Hughes}, Andrew K. and {Motta}, Sara E.},
        title = "{Blast waves and reverse shocks: from ultra-relativistic GRBs to moderately relativistic X-ray binaries}",
      journal = {\mnras},
         year = 2025,
        month = may,
       volume = {539},
       number = {3},
        pages = {2665-2684},
          doi = {10.1093/mnras/staf609},
archivePrefix = {arXiv},
       eprint = {2503.10802},
 primaryClass = {astro-ph.HE},
       adsurl = {https://ui.adsabs.harvard.edu/abs/2025MNRAS.539.2665M}
}

\begin{appendix}

\section{MeerKAT observations}

\begin{table}[!h]
    \caption{Summary of MeerKAT observations of V4641~Sgr.}
    \centering
    \begin{tabular}{lcccc}
    \hline
    Block ID & Date & MJD & Band & Exposure \\
     & & & & [min] \\
    \hline
    1580539555 & 2020-02-01 & 58880.33 & L & 15 \\
    1581132962 & 2020-02-08 & 58887.16 & L & 15 \\
    1634467571 & 2021-10-17 & 59504.49 & L & 15 \\
    1634986873 & 2021-10-23 & 59510.50 & L & 15 \\
    1635685270 & 2021-10-31 & 59518.59 & L & 15 \\
    1636886466 & 2021-11-14 & 59532.49 & L & 15 \\
    1638001320 & 2021-11-27 & 59545.38 & L & 15 \\
    1639226621 & 2021-12-11 & 59559.55 & L & 15 \\
    1639819878 & 2021-12-18 & 59566.43 & L & 15 \\
    1640341567 & 2021-12-24 & 59572.47 & L & 15 \\
    1641196873 & 2022-01-03 & 59582.37 & L & 15 \\
    1641640277 & 2022-01-08 & 59587.50 & L & 15 \\
    1723904179 & 2024-08-17 & 60539.61 & L & 60 \\
    1726494315 & 2024-09-16 & 60569.63 & L & 15 \\
    1726920975 & 2024-09-21 & 60574.57 & L & 15 \\
    1727610136 & 2024-09-29 & 60582.54 & L & 15 \\
    1728213242 & 2024-10-06 & 60589.52 & L & 15 \\
    1728903370 & 2024-10-14 & 60597.53 & L & 15 \\
    1729424896 & 2024-10-20 & 60603.55 & L & 15 \\
    1729943660 & 2024-10-26 & 60609.54 & L & 15 \\
    1730455875 & 2024-11-01 & 60615.46 & L & 15 \\
    1731232274 & 2024-11-10 & 60624.46 & L & 15 \\
    1731749478 & 2024-11-16 & 60630.45 & L & 15 \\
    1732268475 & 2024-11-22 & 60636.46 & L & 15 \\
    1732953016 & 2024-11-30 & 60644.38 & L & 15 \\
    1732960189 & 2024-11-30 & 60644.42 & L & 240 \\
    1733648172 & 2024-12-08 & 60652.43 & L & 15 \\
    1734161776 & 2024-12-14 & 60658.37 & L & 15 \\
    1734764953 & 2024-12-21 & 60665.35 & L & 15 \\
    1735457475 & 2024-12-29 & 60673.37 & L & 15 \\
    1735971973 & 2025-01-04 & 60679.31 & L & 15 \\
    1741576597 & 2025-03-10 & 60744.15 & UHF & 60 \\
    \end{tabular}
    \label{tab:obs_meerkat}
\end{table}

\section{H.E.S.S. and HAWC contours}
\label{appendix:hess_hawc}

\begin{figure*}
    \centering
    \includegraphics[width=\linewidth]{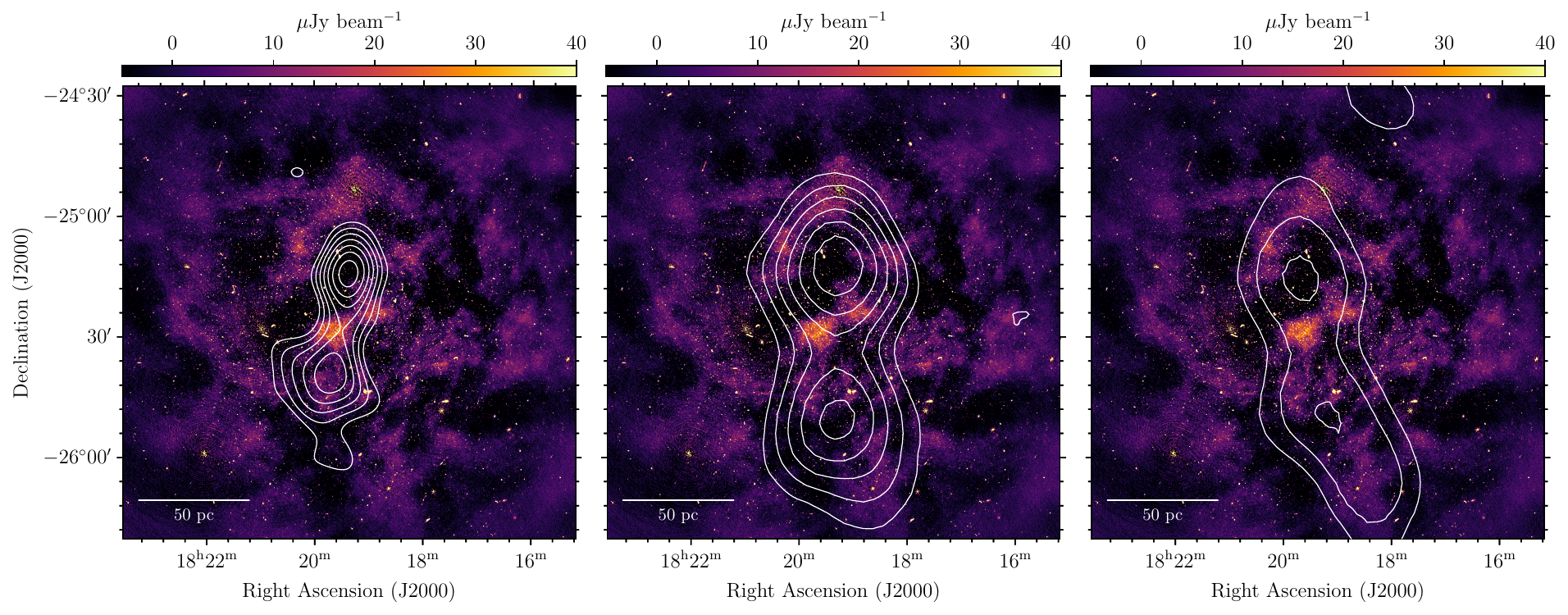}
    \caption{ V4641~Sgr field as observed by the MeerKAT interferometer in the \textit{L} band, along with H.E.S.S. (left) and HAWC (middle: above 1~TeV; right: above 100~TeV) gamma-ray contours at the levels of 3, 4, ..., 9 $\sigma$ \citep{HESS2025, Alfaro2024}.}
    \label{fig:hess_hawc}
\end{figure*}

In Fig. \ref{fig:hess_hawc}, H.E.S.S. and HAWC contours are overlaid on top of the MeerKAT \textit{L}-band image.

\section{Proper motion of nearby stars}
\label{appendix:propermotion}

\begin{figure}
    \centering
    \includegraphics[width=\linewidth]{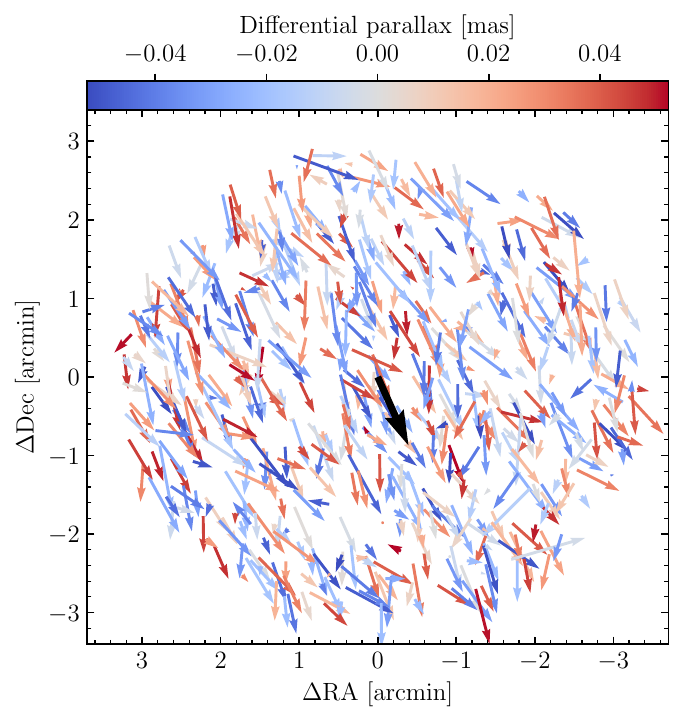}
    \caption{Motion of nearby stars in the plane of the sky, as measured by \textit{Gaia}. Each arrow represent the proper motion (direction and modulus) of a star at a certain position. Only stars within a $3'$ radius and a $p_\star \pm 2\sigma_{p_\star}$ parallax range are shown. The differential parallax between V4641~Sgr and individual stars is color-coded. The black arrow corresponds to the direction of the average proper motion.}
    \label{fig:propermotion_vector}
\end{figure}

Fig. \ref{fig:propermotion_vector} shows another representation of the \textit{Gaia} data displayed in Fig. \ref{fig:propermotion}. Each arrow corresponds to the proper motion of an individual star.

\section{Synchrotron radiation:\ Minimum energy requirements}
\label{appendix:min_energy}

Let us assume that we observe a synchrotron spectrum with some spectral index $\alpha$ ($S_\nu \propto \nu^\alpha$) between two frequencies $\nu_1$ and $\nu_2$. At a distance $D$ the integrated luminosity is then
\begin{equation}
    L = 4\pi D^2 \int_{\nu_1}^{\nu_2} S_\nu \mathrm{d}\nu = 4\pi D^2 S_{\nu_0} \nu_0^{-\alpha} \left[ \frac{\nu_2^{\alpha+1} - \nu_1^{\alpha+1}}{\alpha+1} \right]
,\end{equation}
where $S_{\nu_0}$ is a measurement of the flux density at frequency $\nu_0$.
Within the relativistic plasma the total energy is stored in the magnetic fields and in the electrons. For a source volume $V$, the former is given by
\begin{equation}
    E_B = \frac{B^2}{8\pi} fV
,\end{equation}
where $f$ is a filling factor accounting for the fraction of $V$ which is actually occupied by the relativistic plasma. The energy in electrons, for a power-law distribution of index $p = 2(1-\alpha)$, is given by
\begin{equation}
    E_e = c_{12}(p, \nu_1, \nu_2) B^{-3/2} L
,\end{equation}
where
\begin{equation}
    c_{12}(p, \nu_1, \nu_2) = c_1^{1/2} c_2^{-1} \Tilde{c}(p, \nu_1, \nu_2)
\end{equation}
and
\begin{equation}
    \Tilde{c}(p, \nu_1, \nu_2) = \frac{p-3}{p-2} \left[ \frac{\nu_1^{(2-p)/2} - \nu_2^{(2-p)/2}}{\nu_1^{(3-p)/2} - \nu_2^{(3-p)/2}} \right].
\end{equation}
Since $E_B \propto B^2$ and $E_e \propto B^{-3/2}$, there exists a value of $B$ for which the total energy is minimum. This occurs close to equipartition between energies in electrons and magnetic fields, 
\begin{equation}
    E_B = \frac{3}{4} \eta E_e
,\end{equation}
where $\eta = 1 + \beta$, and $\beta$ is the ratio of energy in protons to that in electrons. For synchrotron sources,  $\eta = 1$ (i.e., ignoring the protons) is generally assumed. This leads to the equipartition magnetic field
\begin{equation}
    B_\mathrm{eq} = \left( \frac{6\pi \eta c_{12} L}{fV} \right)^{2/7}
,\end{equation}
along with the associated total minimum energy,
\begin{equation}
    E_\mathrm{min} = \frac{7}{4} \eta c_{12} L B_\mathrm{eq}^{-3/2}.
\end{equation}

\section{Spectral emissivity of thermal bremsstrahlung radiation}
\label{appendix:free-free}

Consider a plasma of electrons and protons with particle densities $n_e$ and $n$ at temperature $T$. The free-free spectral emissivity is given by
\begin{equation}
    \varepsilon_\nu = \kappa T^{-1/2} n n_e Z^2 g(\nu, T) \exp\left(-\frac{h\nu}{k_B T}\right),
\end{equation}
where
\begin{equation}
    \kappa = \frac{1}{3\pi^2} \left(\frac{\pi}{6}\right)^{1/2} \frac{e^6}{\varepsilon_0^3 c^3 m_e^{3/2} k_B^{1/2}} \simeq 6.84\times 10^{-38}~\mathrm{erg~s^{-1}~Hz^{-1}~cm^3~K^{1/2}}.
\end{equation}
Here, $g(\nu, T)$ is the Gaunt factor and at radio frequencies takes the form of
\begin{equation}
    g(\nu, T) = \frac{\sqrt{3}}{2\pi} \left[ \ln\left(\frac{128\varepsilon_0^2 k_B^3 T^3}{m_e e^4 \nu^2 Z^2}\right) - \gamma^{1/2} \right]
\end{equation}
where $\gamma \simeq 0.577$ is Euler's constant. Assuming a fully ionized hydrogen plasma ($n_e = n$, $Z = 1$), we can infer the electron density from $\varepsilon_\nu$ and $T$ as
\begin{equation}
    n_e = \left[ \frac{\varepsilon_\nu T^{1/2} \exp(h\nu/k_B T)}{\kappa g(\nu, T)} \right]^{1/2}.
\end{equation}

\end{appendix} 

\end{document}